\documentclass[pdflatex,sn-nature,twocolumn,oneside]{sn-jnl}

\usepackage{graphicx}%
\usepackage{multirow}%
\usepackage{amsmath,amssymb,amsfonts}%
\usepackage{amsthm}%
\usepackage{mathrsfs}%
\usepackage[title]{appendix}%
\usepackage{xcolor}%
\usepackage{textcomp}%
\usepackage{manyfoot}%
\usepackage{booktabs}%
\usepackage{algorithm}%
\usepackage{algorithmicx}%
\usepackage{algpseudocode}%
\usepackage{listings}%
\usepackage{cleveref}
\usepackage{chemformula}
\usepackage{siunitx}
\DeclareSIUnit\angstrom{\text{Å}}
\usepackage{subcaption}
\usepackage{orcidlink}
\usepackage{colortbl}
\usepackage{siunitx}

\usepackage{tikz}
\usetikzlibrary{external, matrix}

\crefrangeformat{equation}{Equations~#3#1#4--#5#2#6}
\crefmultiformat{equation}{Equations~#2#1#3}{ and~#2#1#3}{, #2#1#3}{ and~#2#1#3}
\crefformat{equation}{Equation~#2#1#3}
\crefrangeformat{figure}{Figures~#3#1#4--#5#2#6}
\crefmultiformat{figure}{Figures~#2#1#3}{ and~#2#1#3}{, #2#1#3}{ and~#2#1#3}
\crefformat{figure}{Figure~#2#1#3}
\crefrangeformat{table}{Tables~#3#1#4--#5#2#6}
\crefmultiformat{table}{Tables~#2#1#3}{ and~#2#1#3}{, #2#1#3}{ and~#2#1#3}
\crefformat{table}{Table~#2#1#3}
\crefname{section}{Section}{Sections}
\crefname{subsection}{Subsection}{Subsections}
\crefname{appendix}{Appendix}{Appendices}

\usepackage[utf8]{inputenc}
\DeclareUnicodeCharacter{2202}{$\partial$}
\DeclareUnicodeCharacter{03B5}{$\varepsilon$}

\newlength{\onecolwidth}
\newlength{\twocolwidth}
\definecolor{green}{HTML}{8EC291}
\definecolor{teal}{HTML}{34858D}
\definecolor{blue}{HTML}{294079}
\definecolor{red}{HTML}{A90505}
\newcommand{\ket}[1]{|#1\rangle}

\newcommand{\braopket}[3]{\langle#1|#2|#3\rangle}

\begin{document}

\title{Predicting electronic screening for fast Koopmans spectral functional calculations}

\author[1]{\fnm{Yannick} \sur{Schubert}\orcidlink{0000-0002-9676-9134}}

\author[1]{\fnm{Sandra} \sur{Luber}\orcidlink{0000-0002-6203-9379}}

\author[2,3]{\fnm{Nicola} \sur{Marzari}\orcidlink{0000-0002-9764-0199}}

\author*[3,4]{\fnm{Edward} \sur{Linscott}\orcidlink{0000-0002-4967-9873} }\email{edward.linscott@psi.ch}

\affil[1]{\orgdiv{Department of Chemistry}, \orgname{University of Zurich}, \orgaddress{\postcode{8057} \city{Zurich}, \country{Switzerland}}}

\affil[2]{\orgdiv{Theory and Simulations of Materials (THEOS) and National Centre for Computational Design and Discovery of Novel Materials (MARVEL)}, \orgname{École Polytechnique Fédérale de Lausanne}, \orgaddress{\postcode{1015} \city{Lausanne}, \country{Switzerland}}}

\affil[3]{\orgdiv{Center for Scientific Computing, Theory and Data}, \orgname{Paul Scherrer Institute}, \orgaddress{\postcode{5352} \city{Villigen PSI}, \country{Switzerland}}}

\affil[4]{\orgdiv{National Centre for Computational Design and Discovery of Novel Materials (MARVEL)}, \orgname{Paul Scherrer Institute}, \orgaddress{\postcode{5352} \city{Villigen PSI}, \country{Switzerland}}}

\abstract{%
Koopmans spectral functionals are a powerful extension of Kohn-Sham density-functional theory (DFT) that enable the prediction of spectral properties with state-of-the-art accuracy. The success of these functionals relies on capturing the effects of electronic screening through scalar, orbital-dependent parameters. These parameters have to be computed for every calculation, making Koopmans spectral functionals more expensive than their DFT counterparts. In this work, we present a machine-learning model that --- with minimal training --- can predict these screening parameters directly from orbital densities calculated at the DFT level. We show on two prototypical use cases that using the screening parameters predicted by this model, instead of those calculated from linear response, leads to orbital energies that differ by less than 20 meV on average. Since this approach dramatically reduces run-times with minimal loss of accuracy, it will enable the application of Koopmans spectral functionals to classes of problems that previously would have been prohibitively expensive, such as the prediction of temperature-dependent spectral properties. More broadly, this work demonstrates that measuring violations of piecewise linearity (i.e. curvature in total energies with respect to occupancies) can be done efficiently by combining frozen-orbital approximations and machine learning.
}


\maketitle

Predicting the spectral properties of materials from first principles can greatly assist the design of optical and electronic devices \cite{Marzari2021}. Among the various techniques one can employ, Koopmans spectral functionals are a promising approach due to their accuracy and comparably low computational cost \cite{Dabo2009,Dabo2010,Dabo2014,Borghi2014,Ferretti2014,Borghi2015,Nguyen2018,Colonna2018,DeGennaro2022,Colonna2022,Linscott2023}. These functionals are a beyond-DFT extension explicitly designed to predict spectral properties, and have shown success across a range of both isolated and periodic systems. For isolated systems, Koopmans functionals accurately predict the ionization potentials and electron affinities --- and more generally, the orbital energies and photoemission spectra --- of atoms~\cite{Dabo2010}, small molecules~\cite{Borghi2014,Colonna2018,Colonna2019}, organic photovoltaic compounds~\cite{Dabo2013, Nguyen2015}, DNA nucleobases~\cite{Nguyen2016}, and toy models~\cite{Schubert2023}. The method has also been extended to predict optical (i.e. neutral) excitation energies in molecules~\cite{Elliott2019}. For three-dimensional systems, Koopmans functionals have been shown to accurately predict the band structure and band alignment of prototypical semiconductors and insulators \cite{Nguyen2018,DeGennaro2022,Colonna2022}, systems with large spin-orbit coupling \cite{Marrazzo2024}, and a vacancy-ordered double perovskite \cite{Ingall2024}, as well as the band gap of liquid water~\cite{deAlmeida2021}.

One of the crucial quantities involved in the definition of Koopmans functionals is the set of so-called screening parameters, denoted $\{\alpha_i\}$. These parameters account for the effect of electronic screening in the localized basis of the orbitals minimizing the functional. There is one screening parameter per orbital in the system, and each screening parameter can be computed fully from first-principle calculations using finite differences \cite{Dabo2010,Nguyen2018} or with linear-response theory \cite{Colonna2018}. Obtaining reliable screening parameters is essential to the accuracy of Koopmans spectral functionals, while also being the main reason why these functionals are more expensive than their KS-DFT counterparts. The objective of this work is to replace the calculation of screening parameters with a machine-learning (ML) model, thereby drastically reducing the cost of Koopmans functional calculations, and making it possible to apply them more widely.

In the broader context of computational materials science and quantum chemistry, machine learning is being used to predict an ever-increasing range of quantum mechanical properties~\cite{Carleo2019,Ceriotti2021} including electronic excitations~\cite{Dral2021,Westermayr2021,Cignoni2024}. In contrast to many of these approaches, the screening parameters studied in this work are intermediate quantities, not physical observables. In this regard, this work shares parallels with attempts to learn the $U$ parameter for DFT+$U$ functionals \cite{Yu2020,Yu2023,Cai2024,Uhrin2024} and the dielectric screening when solving the Bethe-Salpeter equation~\cite{Dong2021} (albeit in the latter case the dielectric screening is a physical observable, but for the purposes of that work it was used as an ingredient for subsequent calculations of optical spectra). This work also differs from ML methods that seek to relate structural information directly to observable quantities on a practical level~\cite{Montavon2013,Brockherde2017,Welborn2018,Schleder2019,Ryczko2019,Noe2020,Hase2020,Sutton2020,Bogojeski2020,Ghosh2019}, because in our case first-principles calculations will not be bypassed completely.

\begin{figure}[t]
    \centering
    \begin{tikzpicture}
        \matrix[column sep=0cm] {
            \node (a) {\includegraphics[height=0.45\onecolwidth]{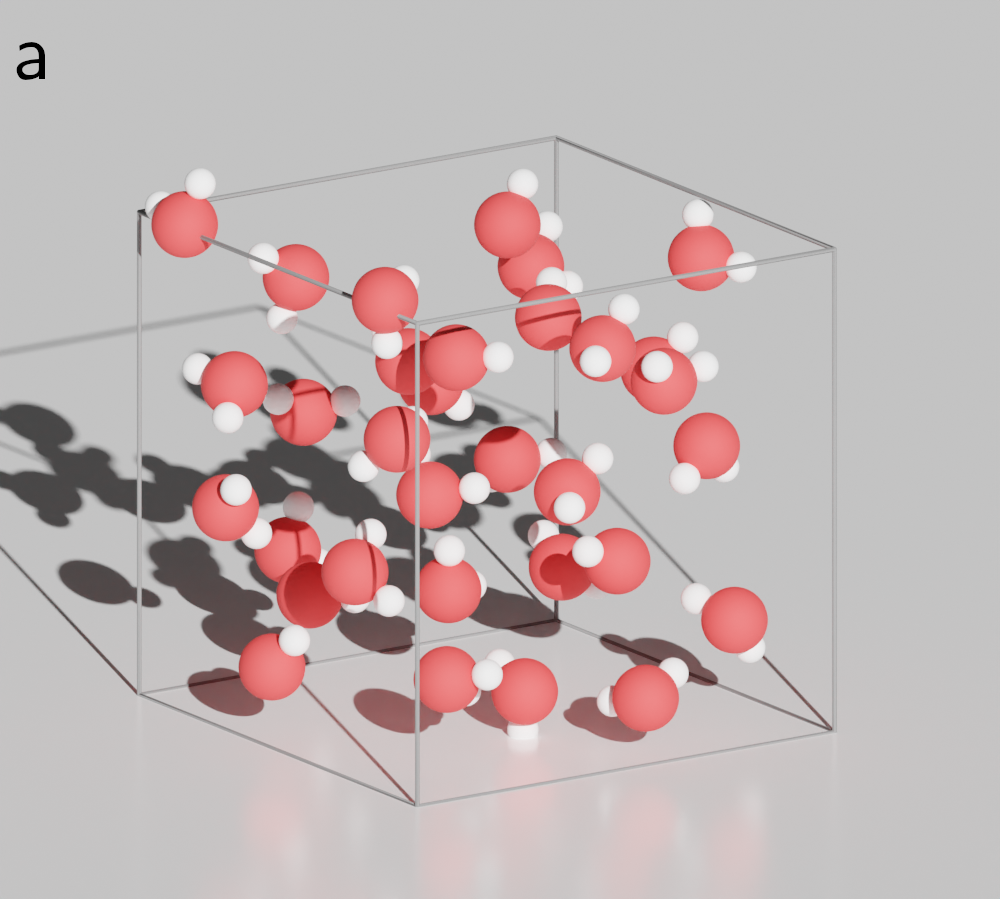}}; &
            \node (b) {\includegraphics[height=0.45\onecolwidth]{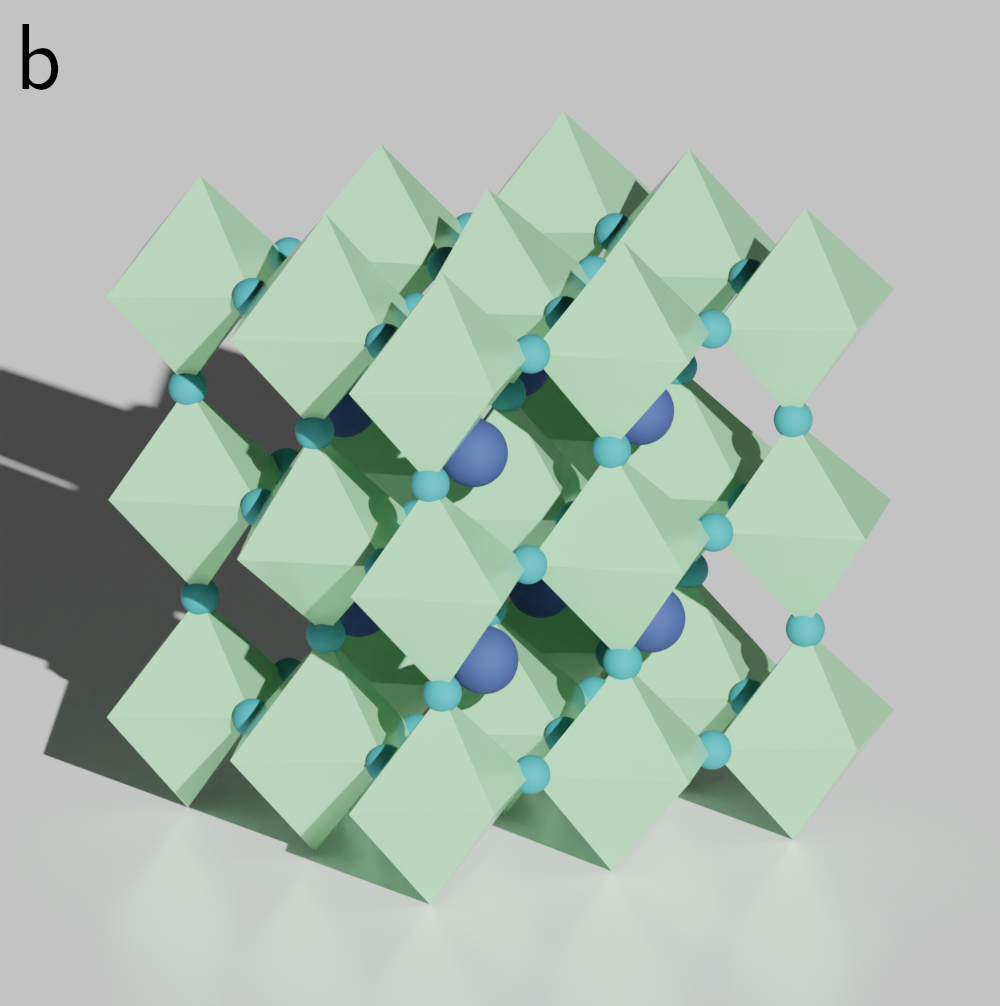}}; \\
        };
    \end{tikzpicture}
    \caption{\textbf{The two systems studied in this work.} (a) Liquid water, with oxygen atoms shown in red and hydrogen in white. (b) The halide perovskite \ch{CsSnI3}, with caesium atoms in blue, tin polyhedra in green, and iodine in teal.}
    \label{fig: systems}
\end{figure}

Specifically, this work presents a simple machine learning framework that can be used to predict the screening parameters for a given chemical system (i.e., we do not develop a general model to predict the screening parameters for an arbitrary chemical system). The two use-cases we will study are liquid water and the halide perovskite \ch{CsSnI3} (\cref{fig: systems}). In the case of liquid water, one might want to calculate spectral properties averaged along a molecular dynamics trajectory. In the case of the halide perovskite, one might want to calculate the temperature-renormalized band structure by performing calculations on an ensemble of structures with the atoms displaced in such a way to appropriately sample the ionic energy landscape \cite{Zacharias2020,Monacelli2021}. In both cases, these methods require Koopmans spectral functional calculations on many copies of the same chemical system, just with different atomic displacements. This process can be made much faster by training a machine learning model on a subset of these copies, and then using this model to predict the screening parameters for the remaining copies.

\section{Koopmans spectral functionals} \label{sec: theory_screening_parameters}
Koopmans spectral functionals are a class of orbital-density dependent functionals that accurately predict spectral properties by imposing the condition that the quasi-particle energies of the functional must match the corresponding total energy difference when an electron is explicitly removed from/added to the system. This quasiparticle/total-energy-difference equivalence is trivially satisfied by the exact one-particle Green's function (as can be seen by the spectral representation, with poles located at energies corresponding to particle addition/removal), but it is violated by standard Kohn-Sham density functionals and leads to, among other failures, semi-local DFT's underestimation of the band gap. Given that the exact Green's function describes one-particle excitations exactly, the idea behind Koopmans functionals is that enforcing this condition on DFT will improve its description of one-particle excitations.

An orbital energy, defined as $\varepsilon_i = \left\langle \varphi_i \right| \hat H \left| \varphi_i \right\rangle$, will be equal to the total energy difference of electron addition/removal if it is independent of that orbital's occupation $f_i$. (This follows because if $\varepsilon_i$ --- which is equal to $\frac{dE}{df_i}$ by Janak's theorem --- is independent of $f_i$, then the total energy difference $\Delta E = {E(f_i = 1)} - {E(f_i = 0)} = \frac{dE}{df_i} = \varepsilon_i$ \cite{Janak1978a}.) Equivalently, the orbital energies will match the corresponding total energy differences if the total energy itself is piecewise-linear with respect to orbital occupations:

\begin{align}
    \frac{d^2E}{df_i^2} =& 0
    \label{eq: GPWL}.
\end{align}

\noindent This condition is referred to as the ``generalised piecewise linearity'' (GPWL) condition. It is related to the more well-known ``piecewise linearity" condition, which states that the exact total energy is piecewise-linear with respect to the \emph{total} number of electrons in the system \cite{Cohen2012}.

Koopmans functionals impose this GPWL condition on a (typically semi-local) DFT functional via tailored corrective terms as follows:

\begin{align}
E^\mathrm{KI} = & E^\mathrm{DFT}[\rho] \nonumber \\
& 
+ \sum_i
\bigg\{
- \left(E^\mathrm{DFT}\big[\rho\big] - E^\mathrm{DFT}\big[\rho^{f_i\rightarrow 0}\big]\right) \nonumber \\
& \hspace{3em}
+ f_i \left(E^\mathrm{DFT}\big[\rho^{f_i\rightarrow 1}\big] - E^\mathrm{DFT}\big[\rho^{f_i \rightarrow 0}\big]\right)
\bigg\} \nonumber \\
=& E^\mathrm{DFT}[\rho] + \sum_i \Pi^\mathrm{KI}_i
\label{eq: koopmans_energy_functional_without_alpha}
\end{align}

\noindent where $\{f_j\}$ are the orbital occupancies and $\rho$ the total electronic density of the $N$-electron ground state, and $\rho^{f_i\rightarrow f}$ denotes the electronic density of the \hbox{($f + \sum_{j\neq i} f_j$)}-electron system with the orbital occupancies constrained to be equal to $\{..., f_{i-1}, f, f_{i+1}, ...\}$ i.e. $\rho^{f_i\rightarrow 1}$ and $\rho^{f_i\rightarrow 0}$ correspond to charged excitations of the ground state where we fill/empty orbital $i$. In the final line of \cref{eq: koopmans_energy_functional_without_alpha} we have introduced the shorthand $\Pi^\mathrm{KI}_i$ for the Koopmans correction to orbital $i$. By construction, this correction removes the non-linear dependence of the DFT energy $E^\mathrm{DFT}$ on the occupation of orbital $i$ (the second line in \cref{eq: koopmans_energy_functional_without_alpha}) and replaces it with a term that is explicitly linear in $f_i$ (the third line), with a slope that corresponds to the finite energy difference between integer occupations of orbital $i$. We note that other choices are possible for the slope of this linear term, which give rise to different variants of Koopmans spectral functionals. In this paper, we will exclusively focus on this ``Koopmans integral'' (KI) variant.

While \cref{eq: koopmans_energy_functional_without_alpha} formally imposes the GPWL condition as desired, practically it cannot be used as a functional because we cannot construct the constrained densities $\rho^{f_i\rightarrow f}$ without explicitly performing constrained DFT calculations.

In order to convert \cref{eq: koopmans_energy_functional_without_alpha} into a tractable form, we instead evaluate these constrained densities in the frozen orbital approximation i.e. we neglect the dependence of the orbitals $\{\varphi_j\}$ on the occupancy of orbital $i$, obtaining

\begin{align}
    \rho^{f_i\rightarrow f}
    \approx 
    f \left|\varphi^N_i\right|^2 + \sum_{j\neq i} f_j \left|\varphi^N_j\right|^2
    = f n_i + \rho - \rho_i
    \label{eq: frozen_orbital_approximation}
\end{align}

\noindent where $\{\varphi^N_j\}$ are the orbitals of the unconstrained $N$-electron ground state and we have introduced the shorthand \hbox{$n_i(\mathbf{r}) = |\varphi^N_i(\mathbf{r})|^2$} for the normalized density of orbital $i$ and $\rho_i(\mathbf{r}) = f_i n_i(\mathbf{r})$. These frozen densities --- unlike their unfrozen counterparts --- are straightforward to evaluate because they are constructed purely from quantities corresponding to the $N$-electron ground state.

Evaluating the KI correction on frozen-orbital densities gives rise to the unscreened KI corrections $\Pi_i^{\mathrm{uKI}}$. The orbital relaxation that is absent in these terms can be accounted for by appropriately screening the Koopmans corrections i.e.

\begin{align}
    \Pi^\mathrm{KI}_i = \alpha_i \Pi^\mathrm{uKI}_i
    \label{eqn: alpha_scaling}
\end{align}

\noindent where $\alpha_i$ is some as-of-yet unknown scalar coefficient that is a measure of how much electronic interactions between orbitals are screened by the rest of the system.  These parameters $\{\alpha_i\}$ are the screening parameters that are the central topic of this work, and they will be discussed in more detail later.

Having introduced the frozen-orbital approximation and compensated for it via the screening parameters, we arrive at the final form of the KI functional:

\begin{align}
    E^\mathrm{KI}_{\boldsymbol\alpha}[\{\rho_j\}] 
    & = E^\mathrm{DFT}[\rho]
    \nonumber \\
    & \hspace{-4em}
    + \sum_i
    \alpha_i
    \bigg\{
    -\left(
    E_\mathrm{Hxc}[\rho]
    - E_\mathrm{Hxc}\big[\rho - \rho_i\big]
    \right)
    \nonumber \\
    & \hspace{-2em}
    + f_i\left(
    E_\mathrm{Hxc}\big[\rho - \rho_i + n_i\big]
    - E_\mathrm{Hxc}\big[\rho - \rho_i\big]
    \right)
    \bigg\}
    \label{eqn: koopmans_energy_functional}.
\end{align}

\noindent where in the Koopmans correction only the Hartree-plus-exchange-correlation term appears because all the other terms in $E^\mathrm{DFT}$ are linear in the orbital occupations and therefore cancel.

In contrast to standard DFT energy functionals, this energy functional is not only dependent on the total density but also on these individual orbital densities $\{\rho_i\}$. Therefore one has to minimize the total energy functional with respect to the entire set of orbital densities $\rho_i$ to obtain the ground state energy, and not just with respect to the total density $\rho$. (This is not so different to Kohn-Sham DFT, where one minimizes the functional with respect to a set of Kohn-Sham orbitals.)

The orbitals $\{\varphi_i\}$ that minimize the Koopmans energy functional are called the variational orbitals. They are found to be localized in space \cite{Ferretti2014, Pederson1984,Pederson1985,Pederson1988,Heaton1983}, closely resembling Boys orbitals in molecules and, equivalently, maximally localized Wannier functions (MLWFs) in solids \cite{Marzari2012}. In the specific case of the KI functional, and unlike most orbital-density-dependent functionals, the total energy is invariant with respect to unitary rotations of the occupied orbital densities. This means that once the variational orbitals are initialized --- typically as MLWFs --- they require no further optimization.

The matrix elements of the KI Hamiltonian are given by

\begin{align}
    \lambda_{ij}^\mathrm{KI}= \langle \varphi_i |\hat H^\mathrm{KI} | \varphi_j \rangle = \langle \varphi_i| \hat{h}^\mathrm{DFT}+\alpha_j\hat{v}_j^\mathrm{KI}|\varphi_j\rangle, \label{eq: lambda_ij Koopmans}
\end{align}

\noindent where the orbital-dependent Koopmans potential $\hat v_j^\mathrm{KI}$ is given by a derivative of the unscreened KI correction $\Pi^\mathrm{uKI}_i$, and is discussed in detail in Ref.~\citenum{Borghi2014}. At the energy minimum, the matrix $\lambda_{ij}^\mathrm{KI}$ becomes Hermitian
\cite{Stengel2008,Borghi2015,Goedecker1997}
and can be diagonalized. The corresponding eigenfunctions are called canonical orbitals and the eigenvalues $\varepsilon_j^\mathrm{KI}$ canonical energies. These canonical orbitals are different from the variational orbitals, but they are related via a unitary transformation, and both give rise to the same total density. Contrast this with KS-DFT functionals, which are invariant with respect to unitary rotations of the set of occupied Kohn-Sham orbitals, and thus the same orbitals both minimize the total energy and diagonalize the Hamiltonian.

Koopmans spectral functionals follow the widely-adopted approach of interpreting canonical orbitals as Dyson orbitals and their energies as quasi-particle energies \cite{Pederson1984,Korzdorfer2008,Vydrov2005,Ortiz2020}.

Before proceeding, it should be noted that while the Koopmans correction is designed to impose an exact condition, the correction itself it is not formally derived from an exact theory such as generalized Kohn-Sham theory (\emph{cf.} hybrid functionals \cite{Garrick2020}). In this sense, Koopmans functionals are not fully \emph{ab initio}. Work is ongoing to provide a more formally rigorous basis for the Koopmans correction. It should also be noted that Koopmans functionals share many similarities with other methods that use the concept of piecewise linearity to either parametrize or correct density functionals. These include DFT+\emph{U} \cite{Anisimov1997a,Dudarev1998a,Pickett1998a,Cococcioni2005a} and its various extensions \cite{Campo2010a,Bajaj2017,Burgess2023}, optimally-tuned hybrid functionals \cite{Stein2010,Kronik2012,Wing2021}, global and localized orbital scaling corrections \cite{Zheng2011,Li2015b,Zheng2015,Li2018,Mei2020,Yang2020,Mahler2022a}, and the Wannier-Koopmans method \cite{Ma2016,Weng2017,Weng2020}. There are also connections between Koopmans functionals and reduced density matrix functional theory \cite{Pernal2005,Pernal2016}, the screened extended Koopmans' theorem \cite{DiSabatino2023}, and ensemble density functional theory \cite{Cernatic2022}. We will return to discuss how the results of this work might be applied to these related methods.

\subsection{Screening parameters}

Let us now return to the screening parameters $\{\alpha_i\}$ that were introduced in the previous section without explaining what they are nor how one might compute them.

As per their definition in \cref{eqn: alpha_scaling}, the screening parameters are responsible for relating the unscreened Koopmans correction $\Pi_i^{\mathrm{uKI}}$ (which we can directly evaluate) to the screened Koopmans correction $\Pi_i^{\mathrm{KI}}$ (which we cannot). The degree to which these two quantities differ will depend on how accurate the approximation of \cref{eq: frozen_orbital_approximation} is i.e. how closely the frozen density $f n_i + \rho - \rho_i$ (where the occupation of orbital $i$ is set to $f$ and all the other orbitals are frozen in the $N$-electron solution) matches the true constrained density $\rho^{f_i\rightarrow f}$ (where all the orbitals are allowed to change in order to minimize the total energy of the $(N - f_i + f)$-electron system). Consequently, the $\alpha_i \rightarrow 1$ limit corresponds to an orbital that if its occupation changes, the rest of the electronic density does not respond and remains totally unchanged i.e. the frozen density matches the true constrained density. This limit arises when orbital $i$ interacts very weakly with the rest of the system/the permittivity is very small. In contrast, the $\alpha_i \rightarrow 0$ limit corresponds an orbital that if its occupation changes, the rest of the system will screen the change in the density: the other orbital densities will change such that the total electronic density remains unchanged i.e. $\rho^{f_i \rightarrow f} \rightarrow \rho$ and thus $\Pi_i^\mathrm{KI} \rightarrow 0$. This limit arises when orbital $i$ interacts very strongly with the rest of the system/the permittivity is very large.

But how might we calculate the screening parameters? They cannot be system-agnostic (\emph{\`a la} mixing parameters in typical hybrid functionals), because they depend on the screening of electronic interactions between orbital densities, which will clearly change from one orbital to the next, let alone one system to the next. To obtain system-specific screening parameters, one can calculate them \emph{ab initio} by finding the value that guarantees that the generalized piecewise linearity condition (\cref{eq: GPWL}) is satisfied. It can be shown that for the KI functional, the screening parameter $\alpha_i$ that will satisfy the GPWL condition for orbital $i$ is given by

\begin{equation}
    \alpha_i =\alpha_i^{0} \frac{\Delta {E_i^\mathrm{DFT}}-{\lambda_{ii}^\text{DFT}}}{\lambda_{ii}^\mathrm{KI}[\{\alpha_i^{0}\}]-{\lambda_{ii}^\text{DFT}}} \label{eq: update alpha},
\end{equation}

\noindent where $\lambda_{ii}^\mathrm{KI}[\{\alpha_i^{0}\}]$ is the $i$\textsuperscript{th} diagonal element of the KI Hamiltonian matrix obtained with an initial guess $\{\alpha_i^{0}\}$ for the screening parameters, and $\lambda_{ii}^\text{DFT}$ is the diagonal matrix element corresponding to orbital $i$ of the base DFT Hamiltonian. $\Delta {E_i^\mathrm{DFT}}$ is the difference in total energy for a charge-neutral calculation and a constrained DFT calculation where orbital $i$ is explicitly emptied \cite{Dabo2014,Nguyen2018}. (\Cref{eq: update alpha} only applies to occupied orbitals; an analogous formulation exists for unoccupied orbitals.) If we take \cref{eq: update alpha} to second order, then the screening coefficients become

\begin{equation}
\alpha_i = \frac{\langle n_i | \epsilon^{-1} f_\mathrm{Hxc}| n_i \rangle}{\langle n_i | f_\mathrm{Hxc}| n_i \rangle}
\label{eq: alpha_LR}
\end{equation}

\noindent where $f_\mathrm{Hxc}$ is the Hartree-plus-exchange-correlation kernel and $\epsilon$ is the non-local microscopic dielectric function \emph{i.e.} the screening parameters are an orbital-resolved measure of how much the electronic interactions are screened by the rest of the system. \Cref{eq: alpha_LR} can be evaluated using density-functional perturbation theory \cite{Colonna2018,Colonna2022}.

Thus, to obtain the screening parameters \emph{ab initio} for a given system one must evaluate either \cref{eq: update alpha} (via finite-difference calculations \cite{Nguyen2018,DeGennaro2022}) or \cref{eq: alpha_LR} (via density-functional perturbation theory \cite{Colonna2022}). For the former \cref{eq: update alpha} must in principle be solved iteratively, to account for the dependence of the variational orbitals on $\{\alpha_i\}$. However, for the particular case of the KI functional, the occupied orbitals are independent of the screening parameters (this is not true for the empty orbitals in theory, but in practice the dependence of these orbitals on the screening parameters is sufficiently weak that it can be neglected). We stress that in this approach the screening parameters are \emph{not} fitting parameters. They are determined via a series of DFT calculations, and are not adjusted to fit experimental data, nor results from higher-order computational methods.

Given that (a) Koopmans spectral functionals are orbital-density-dependent and (b) one must compute the set of screening parameters, a Koopmans functional calculation involves a few additional steps compared to a typical semi-local DFT calculation. In brief, the procedure for calculating quasi-particle energies for periodic systems using the KI functional involves the following four-step workflow:
\begin{enumerate}
    \item an initial Kohn-Sham DFT calculation is performed to obtain the ground-state density;
    \item a Wannierization of the DFT ground state to obtain a set of localized orbitals that are used to initialize/define the variational orbitals $\{\varphi_i\}$ \cite{Marzari2012};
    \item the screening parameters for the variational orbitals are calculated via a series of DFT (or DFPT) calculations;
    \item the KI Hamiltonian 
        \begin{align}
        \lambda_{ij}^\mathrm{KI}=\langle \varphi_i| \hat{h}^\mathrm{DFT}+\alpha_j\hat{v}_j^\mathrm{KI}|\varphi_j\rangle \label{eq: lambda_ij KI}
        \end{align}%
is constructed and diagonalized to obtain the canonical eigenenergies $\varepsilon_i^\text{KI}$.
\end{enumerate}
The third step --- that is, calculating the screening parameters --- typically dominates the computational cost of the workflow. This is because one has to compute one screening parameter per variational orbital, and for each orbital one must perform a finite difference or DFPT calculation. Of course, if two variational orbitals are symmetrically equivalent, then the same screening parameter can be used for both, and thus for highly symmetric, ordered systems the cost of calculating the screening parameters can be substantially reduced. Nevertheless, for large and/or disordered systems the computation of the screening parameters remains by far the most expensive step of the entire Koopmans workflow.

\begin{figure*}[t]
    \includegraphics[width=\twocolwidth]{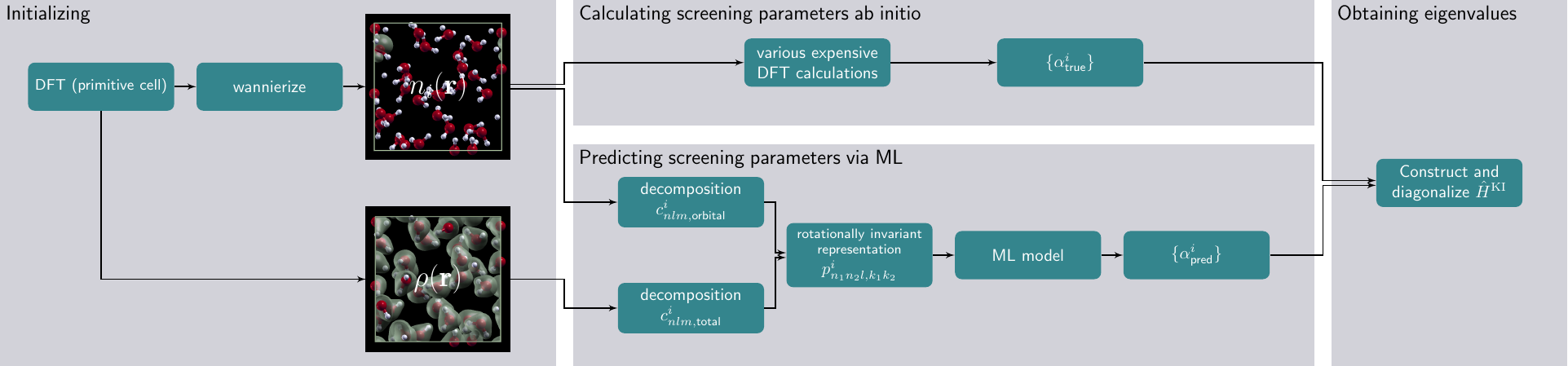}
    \caption{\textbf{Two possible workflows for performing Koopmans spectral functional calculations.} The screening parameters can either be computed \emph{ab initio} (upper pathway) or by a ML model based on the orbital densities $n_i(\mathbf{r})$ and the total density $\rho(\mathbf{r})$ via a rotationally- and translationally-invariant power spectrum descriptor (lower pathway).}
    \label{fig: overview workflow}
\end{figure*}

\section{The machine learning model} \label{sec: ml design}
To accelerate Koopmans spectral functional calculations, this work introduces a machine learning (ML) model to predict the screening parameters, thereby avoiding the expensive step of having to calculate them explicitly. Since this ML model only predicts the screening parameters and not the quasi-particle energies directly, the initial DFT and Wannier calculations are still required to define the variational orbitals, as is the final calculation to calculate the quasi-particle energies. This workflow is compared against a conventional Koopmans functional calculation in \cref{fig: overview workflow}.

\subsection{The descriptors}
In order to design a machine learning model, the first thing to do is to define a set of descriptors, for which there are many possible choices. A preliminary study on acenes (presented in Supplementary Discussion~1) suggested that there is a strong correlation between the self-Hartree energy of each orbital

\begin{equation}
    E_\mathrm{sH}[n_i]=\frac{1}{2}\int \frac{n_i(\textbf{r})n_i(\textbf{r}')}{|\textbf{r}-\textbf{r}'|}d\textbf{r}d\textbf{r}' \label{eq: sH}
\end{equation}

\noindent and its corresponding screening parameter. This scalar quantity is a measure of how localized an orbital density is, and the fact that the screening parameters were correlated with the self-Hartree energies indicated that it might be possible to predict the screening parameters based on a more complete descriptor of the orbital density.

To convert a real-space orbital density $n_i(\mathbf{r})$ into a compact but information-dense descriptor, we define a local decomposition of the normalized orbital density around the orbital's center $\textbf{R}^i$:

\begin{equation}
    c^i_{nlm,k=\mathrm{orbital}}=\int d\textbf{r} \, g_{nl}(r)Y_{lm}(\theta,\varphi)n^i(\textbf{r}-\textbf{R}^i)
\end{equation}

\noindent where $g_{nl}(r)$ are Gaussian radial basis functions and $Y_{lm}(\theta,\varphi)$ are real-valued spherical harmonics (following the choice of Ref.~\citenum{Himanen2020} and many others). For more details regarding these basis functions, refer to Supplementary Discussion~2.

This choice of basis functions means that the model has four hyperparameters: a maximum order for the radial basis functions $n_\mathrm{max}$, a maximum angular orbital momentum $l_\mathrm{max}$, and two radii $r_\mathrm{min}$ and $r_\mathrm{max}$ that quantify the radial extent of the Gaussian basis functions. By design, these basis functions will capture orbital densities most accurately in the vicinity of the orbital's center $\mathbf{R}^i$, and progressively less accurately at larger radii and higher angular momenta. Because the variational orbitals are localized, it is reasonable to assume that most of the relevant information is captured with this local expansion. Choosing these hyperparameters will be influenced by the expected degree of transferability: too large descriptors will lead to more trainable weights and hence usually require more training data. On the other hand, too small descriptors might not capture the information required to accurately predict screening parameters.

In addition to the orbital density $n_i(\mathbf{r})$, one expects from physical intuition that the screening parameter of orbital $i$, $\alpha_i$, will also depend on the surrounding electronic density, because these electrons will also contribute to the local electronic screening. For this reason we construct an analogous local descriptor of the \emph{total} electronic density around the orbital's center $\textbf{R}^i$:

\begin{equation}
    c^i_{nlm,k=\mathrm{total}}=\int d\textbf{r} g_{nl}(r)Y_{lm}(\theta,\varphi)\rho(\textbf{r}-\textbf{R}^i).
\end{equation}

\noindent Finally, it is important to ensure that the descriptors are invariant with respect to translations and rotations of the entire system, so that the network does not consume training data learning that these operations will not affect the screening parameters. The coefficient vectors $\{\mathbf{c}^i\}$ defined above are already invariant with respect to translation, but not with respect to rotation. To obtain rotationally invariant input vectors we construct a power spectrum from these coefficient vectors, explicitly coupling coefficients belonging to different shells $n$, as well as coefficients belonging to the total and to the orbital density. The resulting input vector $\textbf{p}^i$ for each orbital $i$ is given by the vector containing the coefficients

\begin{align*}
    p^i_{n_1n_2l,k_1k_2}=\pi \sqrt{\frac{8}{2l+1}}\sum\limits_m {c_{n_1lm,k_1}^{i *}}c_{n_2lm,k_2}^i 
\end{align*}

\noindent corresponding to all possible combinations of $n_1$, $n_2$, $k_1$ and $k_2$~\cite{Bartok2013}.

\subsection{The network}
Having defined a descriptor, we must now decide on a machine learning model with which to map the power spectrum of each orbital to its screening parameter. In this work, we use ridge regression~\cite{Hoerl1970}. Despite its simplicity, we found that it achieved sufficient accuracy for the case studies with very little training data. By contrast, complex neural networks have more trainable parameters and therefore typically would require more training data. (Of course, this work does not preclude the possibility of employing more sophisticated models in the future.)

\section{Results}\label{sec: results}

\subsection{Test systems}
This machine learning framework was tested on two systems: liquid water and the halide perovskite \ch{CsSnI3}.

Despite its simple molecular structure, water exhibits very complex behavior. Understanding it better would help us to improve our ability to explain and predict a variety of phenomena in nature and technology and it is therefore an active area of research \cite{deAlmeida2021, Gaiduk2018}. For example, accurate values for the ionization potential (IP) and the electron affinity (EA) of water are necessary for a precise description of redox reactions in aqueous systems. These, in turn, are key to many applications such as (photo-)electrochemical cells.

Perovskite solar cells, meanwhile, are one of the most promising candidates for next-generation solar cells, with reported efficiencies higher than conventional silicon-based solar cells \cite{Min2021}. One of the most prominent perovskite materials for solar cell applications is caesium lead halide (\ch{CsPbI3}) due to its suitable band gap of $\SI{1.73}{eV}$ and its excellent electronic properties. The main drawback of \ch{CsPbI3} is that lead is toxic, and it is desirable to find more environmentally friendly metals whose substitution does not compromise performance \cite{Li2020}. \ch{CsSnI3} is one such candidate.

Koopmans calculations were performed on 20 uncorrelated snapshots (\emph{i.e.} different copies of the system with different atomic geometries) for each of the two test systems. Screening parameters were computed \emph{ab initio} for all 20 snapshots; the first 10 snapshots were used when training the models, and the remaining 10 were exclusively reserved for validation. Further details can be found in the Methods section. The water test case follows what one could do in order to calculate the spectral properties of water with Koopmans functionals, where one must average across a molecular dynamics trajectory, while the perovskite test case represents how one could calculate the temperature-dependence of the spectral properties of this system, accounting for anharmonic nuclear motion.

\subsection{Accuracy}
To evaluate the accuracy of a model that predicts screening parameters, we can examine several different metrics. The most obvious quantities to compare are the predicted and the calculated screening parameters. However, these parameters are not physical observables: they are intermediate parameters internal to the Koopmans spectral functional framework. Ultimately, it is much more important that the canonical eigenenergies are accurately predicted. This is because all spectral properties derive from the eigenenergies, and spectral properties are of central interest whenever Koopmans spectral functionals are used.

That said, the eigenenergies are closely related to the screening parameters. As described previously, the eigenenergies are the eigenvalues of the matrix $\lambda_{ij}^\mathrm{KI}$, whose elements contain the screening parameters (\cref{eq: lambda_ij KI}). If all of the orbitals had the same screening parameter $\alpha$ then the difference between the Koopmans and DFT eigenvalues would be linear in $\alpha$. For a system with non-uniform screening parameters, the relationship between the eigenvalues and the screening parameters is more complex, with the difference between the Koopmans and DFT eigenvalues becoming a linear mix of the screening parameters of variational orbitals that constitute the canonical orbital in question. More concretely, given that the variational and canonical orbitals are related via a unitary rotation (i.e. $\ket{\psi_i} = \sum_j U_{ij} \ket{\varphi_j})$, it follows that the Koopmans correction shifts DFT quasi-particle energies by

\begin{align}
    \Delta \varepsilon_i = \varepsilon^\mathrm{KI}_i - \varepsilon^\mathrm{DFT}_i
    =
    \sum_{jk} \alpha_j U_{ij}U_{ki}^\dag\braopket{\varphi_k}{\hat v_j^\mathrm{KI}}{\varphi_j}
\end{align}

\noindent which is proportional to $\alpha_j$, with a constant of proportionality corresponding to the degree of overlap between canonical-variational orbital pairs, as well as Koopmans potential matrix elements. In general the matrix element $\braopket{\varphi_k}{\hat v_j^\mathrm{KI}}{\varphi_j}$ is non-diagonal and non-local, but for occupied orbitals it is diagonal and scalar, and the above expression simplifies to

\begin{align}
    \Delta \varepsilon_{i \in \mathrm{occ}} = 
    \sum_{j} \alpha_j U_{ij}U_{ji}^\dag
    \bigg( & -E_\mathrm{Hxc}[\rho - n_j]+E_\mathrm{Hxc}[\rho]
    \nonumber \\
    & - \int d\mathbf{r} \, v_\mathrm{Hxc}[\rho](\mathbf{r})  n_j(\mathbf{r}) \bigg)
\end{align}

\noindent For a quantitative example, sweeping the screening parameters across the range of \emph{ab initio} values observed in liquid water changes the eigenvalues by approximately 1 eV (see Supplementary Discussion~3).

\begin{figure}[t]
    \centering
    \includegraphics[width=\onecolwidth]{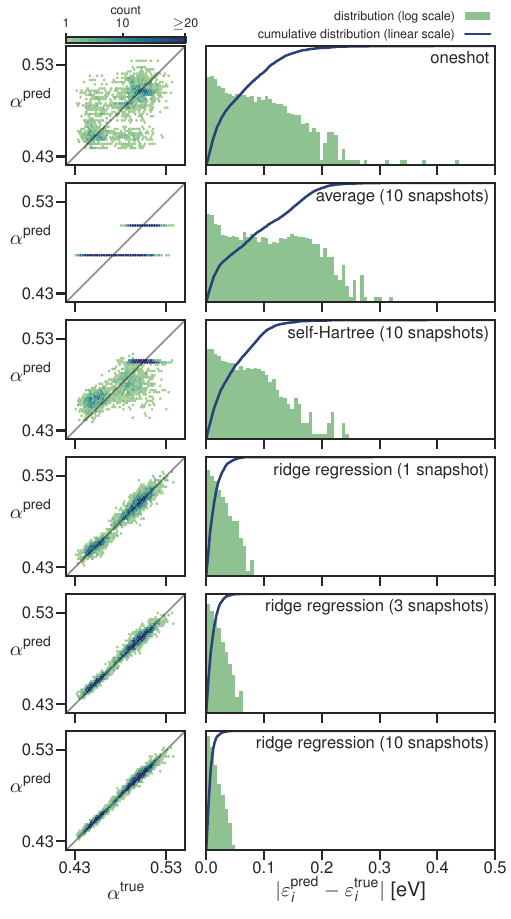}
    \caption{\textbf{Accuracy of the predicted screening parameters and KI@PBE eigenvalues for water with six different models.} The left-hand panels compare the predicted screening parameters against those computed \emph{ab initio} (binned and colored by frequency). The right panels, meanwhile, show logarithmically-scaled error histograms of the absolute error in the eigenvalues obtained using predicted screening parameters compared to eigenvalues obtained using the \emph{ab initio} screening parameters. The blue lines show the corresponding cumulative distributions of these absolute errors.}
    \label{fig: water_cls}
\end{figure}

\begin{figure}[t]
    \centering
    \includegraphics[width=\onecolwidth]{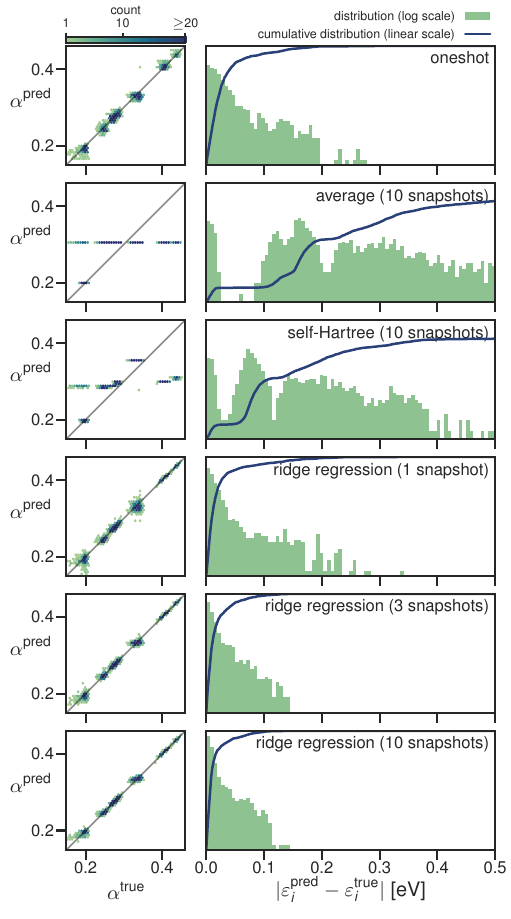}
    \caption{\textbf{Accuracy of the prediction of the screening parameters and KI@PBEsol eigenvalues for \ch{CsSnI3} with six different models.} See \cref{fig: water_cls} for further explanation.}
    \label{fig: CsSnI3}
\end{figure}

The accuracy of the predicted screening parameters and eigenenergies are shown in \cref{fig: water_cls} for water and in \cref{fig: CsSnI3} for \ch{CsSnI3}. In these figures, we compare the performance of the ML model against two simplistic benchmark models:
\begin{itemize}
    \item the ``oneshot'' model, in which the screening parameters are computed \emph{ab initio} for one snapshot, and then this set of screening parameters is used on other snapshots (i.e. neglecting the dependence of the screening parameters on the atomic positions)
    \item the ``average'' model, which takes the average of the \emph{ab initio} screening parameters of the training snapshots as the prediction for all screening parameters of the testing snapshots.
\end{itemize}
Given an arbitrary training dataset, these are the simplest possible models for predicting screening parameters. Any successful alternative model must therefore substantially outperform them, and as such they serve as useful benchmarks.

We will also compare against a ``self-Hartree'' (sH) model; a linear regression model with the self-Hartree energies of each orbital (\cref{eq: sH}) as input and the screening parameters of each orbital as output (inspired by the preliminary study discussed in Supplementary Discussion~1). Both the average and self-Hartree models treated occupied and empty states separately because this gave better results than using one model for the occupied and the empty states together. 
We note that this treatment of empty states is only possible because the empty states are localized and thus their self-Hartree energies are well-defined. (This was not the case for the preliminary study on acenes presented in the Supplementary Information.)

\cref{fig: water_cls,fig: CsSnI3} shows the accuracy of the screening parameters and eigenvalues predicted by six different models: the oneshot model, the average model (trained on 10 snapshots), the self-Hartree model (trained on 10 snapshots), and the ridge-regression model (trained on 1, 3, and 10 snapshots). The accuracy of all six models was assessed against 10 unseen test snapshots. The mean and maximum absolute errors in the eigenvalues are also tabulated in \cref{tab: eigenvalue_errors}.

\begin{table*}[t]
    \centering
    \caption{The mean and maximum absolute error (in meV) of the quasiparticle energies calculated using screening parameters from different models relative to those calculated using \emph{ab initio} screening parameters across the 10 unseen test snapshots. The table presents the mean and maximum absolute errors in all the eigenvalues, in the valence band maxima (VBM), in the conduction band minima (CBM), and in the band gaps $E_g$.}
    \scriptsize
    \renewcommand{\arraystretch}{1.6}
    \begin{tabular}{l r S[table-format=3.1] S[table-format=3.1] *{3}{c S[table-format=3.1] S[table-format=3.1]}}
    & & \multicolumn{11}{c}{\bf water} \\
    & & \multicolumn{2}{c}{all} & & \multicolumn{2}{c}{VBM} & & \multicolumn{2}{c}{CBM} & & \multicolumn{2}{c}{band gap} \\
    & & \multicolumn{1}{c}{mean} & \multicolumn{1}{c}{max} & & \multicolumn{1}{c}{mean} & \multicolumn{1}{c}{max} & & 
    \multicolumn{1}{c}{mean} & \multicolumn{1}{c}{max} & & \multicolumn{1}{c}{mean} & \multicolumn{1}{c}{max} \\
    \multicolumn{2}{c}{oneshot} & 
{\cellcolor[HTML]{EEF8A4}} \color[HTML]{000000} 58.0 & {\cellcolor[HTML]{D43D4F}} \color[HTML]{F1F1F1} 436.6 & & {\cellcolor[HTML]{FECA79}} \color[HTML]{000000} 121.9 & {\cellcolor[HTML]{D43D4F}} \color[HTML]{F1F1F1} 256.1 & & {\cellcolor[HTML]{91D3A4}} \color[HTML]{000000} 7.9 & {\cellcolor[HTML]{A7DBA4}} \color[HTML]{000000} 16.0 & & {\cellcolor[HTML]{FECC7B}} \color[HTML]{000000} 120.4 & {\cellcolor[HTML]{D43D4F}} \color[HTML]{F1F1F1} 267.5 \\
    \multicolumn{2}{c}{average} & 
{\cellcolor[HTML]{FFFCBA}} \color[HTML]{000000} 81.3 & {\cellcolor[HTML]{D43D4F}} \color[HTML]{F1F1F1} 321.5 & & {\cellcolor[HTML]{FDAD60}} \color[HTML]{000000} 140.2 & {\cellcolor[HTML]{D43D4F}} \color[HTML]{F1F1F1} 216.0 & & {\cellcolor[HTML]{97D5A4}} \color[HTML]{000000} 8.9 & {\cellcolor[HTML]{A2D9A4}} \color[HTML]{000000} 14.7 & & {\cellcolor[HTML]{FDB768}} \color[HTML]{000000} 133.7 & {\cellcolor[HTML]{D43D4F}} \color[HTML]{F1F1F1} 213.3 \\
    \multicolumn{2}{c}{self-Hartree} & 
{\cellcolor[HTML]{E1F399}} \color[HTML]{000000} 46.7 & {\cellcolor[HTML]{D43D4F}} \color[HTML]{F1F1F1} 241.1 & & {\cellcolor[HTML]{E7F59A}} \color[HTML]{000000} 49.2 & {\cellcolor[HTML]{FEEB9D}} \color[HTML]{000000} 98.3 & & {\cellcolor[HTML]{97D5A4}} \color[HTML]{000000} 9.0 & {\cellcolor[HTML]{A4DAA4}} \color[HTML]{000000} 15.5 & & {\cellcolor[HTML]{DFF299}} \color[HTML]{000000} 45.1 & {\cellcolor[HTML]{FEE593}} \color[HTML]{000000} 103.4 \\
    \multirow[c]{3}{*}{ridge regression} & 1 &
{\cellcolor[HTML]{A2D9A4}} \color[HTML]{000000} 13.6 & {\cellcolor[HTML]{FFFCBA}} \color[HTML]{000000} 81.8 & & {\cellcolor[HTML]{BFE5A0}} \color[HTML]{000000} 28.6 & {\cellcolor[HTML]{F1F9A9}} \color[HTML]{000000} 60.8 & & {\cellcolor[HTML]{8CD1A4}} \color[HTML]{000000} 5.7 & {\cellcolor[HTML]{9FD8A4}} \color[HTML]{000000} 12.9 & & {\cellcolor[HTML]{C1E6A0}} \color[HTML]{000000} 30.2 & {\cellcolor[HTML]{EDF8A3}} \color[HTML]{000000} 56.9 \\
    & 3 & 
{\cellcolor[HTML]{97D5A4}} \color[HTML]{000000} 10.3 & {\cellcolor[HTML]{F1F9A9}} \color[HTML]{000000} 61.9 & & {\cellcolor[HTML]{A4DAA4}} \color[HTML]{000000} 15.5 & {\cellcolor[HTML]{DDF19A}} \color[HTML]{000000} 43.3 & & {\cellcolor[HTML]{86CFA5}} \color[HTML]{000000} 3.0 & {\cellcolor[HTML]{94D4A4}} \color[HTML]{000000} 8.7 & & {\cellcolor[HTML]{A7DBA4}} \color[HTML]{000000} 16.9 & {\cellcolor[HTML]{E1F399}} \color[HTML]{000000} 46.5 \\
    & 10\vspace{0.5em}& 
{\cellcolor[HTML]{91D3A4}} \color[HTML]{000000} 6.9 & {\cellcolor[HTML]{DFF299}} \color[HTML]{000000} 45.0 & & {\cellcolor[HTML]{97D5A4}} \color[HTML]{000000} 10.3 & {\cellcolor[HTML]{C1E6A0}} \color[HTML]{000000} 29.7 & & {\cellcolor[HTML]{86CFA5}} \color[HTML]{000000} 2.7 & {\cellcolor[HTML]{8FD2A4}} \color[HTML]{000000} 6.3 & & {\cellcolor[HTML]{99D6A4}} \color[HTML]{000000} 10.8 & {\cellcolor[HTML]{BCE4A0}} \color[HTML]{000000} 27.3 \\
& & \multicolumn{11}{c}{\bf CsSnI\textsubscript{3}} \\
    \multicolumn{2}{c}{oneshot} & 
{\cellcolor[HTML]{BCE4A0}} \color[HTML]{000000} 29.8 & {\cellcolor[HTML]{D43D4F}} \color[HTML]{F1F1F1} 272.3 & & {\cellcolor[HTML]{B8E2A1}} \color[HTML]{000000} 28.2 & {\cellcolor[HTML]{E1F399}} \color[HTML]{000000} 49.2 & & {\cellcolor[HTML]{A7DBA4}} \color[HTML]{000000} 20.2 & {\cellcolor[HTML]{BCE4A0}} \color[HTML]{000000} 30.9 & & {\cellcolor[HTML]{E1F399}} \color[HTML]{000000} 48.2 & {\cellcolor[HTML]{FFFFBE}} \color[HTML]{000000} 80.1 \\
    \multicolumn{2}{c}{average} & 
{\cellcolor[HTML]{D43D4F}} \color[HTML]{F1F1F1} 263.1 & {\cellcolor[HTML]{D43D4F}} \color[HTML]{F1F1F1} 946.8 & & {\cellcolor[HTML]{D43D4F}} \color[HTML]{F1F1F1} 374.6 & {\cellcolor[HTML]{D43D4F}} \color[HTML]{F1F1F1} 404.2 & & {\cellcolor[HTML]{89D0A4}} \color[HTML]{000000} 7.0 & {\cellcolor[HTML]{97D5A4}} \color[HTML]{000000} 12.5 & & {\cellcolor[HTML]{D43D4F}} \color[HTML]{F1F1F1} 372.5 & {\cellcolor[HTML]{D43D4F}} \color[HTML]{F1F1F1} 409.9 \\
    \multicolumn{2}{c}{self-Hartree} & 
{\cellcolor[HTML]{D43D4F}} \color[HTML]{F1F1F1} 209.1 & {\cellcolor[HTML]{D43D4F}} \color[HTML]{F1F1F1} 934.8 & & {\cellcolor[HTML]{D43D4F}} \color[HTML]{F1F1F1} 285.6 & {\cellcolor[HTML]{D43D4F}} \color[HTML]{F1F1F1} 315.6 & & {\cellcolor[HTML]{86CFA5}} \color[HTML]{000000} 6.1 & {\cellcolor[HTML]{94D4A4}} \color[HTML]{000000} 11.9 & & {\cellcolor[HTML]{D43D4F}} \color[HTML]{F1F1F1} 284.7 & {\cellcolor[HTML]{D43D4F}} \color[HTML]{F1F1F1} 322.9 \\
    \multirow[c]{3}{*}{ridge regression} & 1 &
{\cellcolor[HTML]{AEDEA3}} \color[HTML]{000000} 23.1 & {\cellcolor[HTML]{D43D4F}} \color[HTML]{F1F1F1} 339.4 & & {\cellcolor[HTML]{A7DBA4}} \color[HTML]{000000} 19.8 & {\cellcolor[HTML]{CAEA9E}} \color[HTML]{000000} 38.0 & & {\cellcolor[HTML]{A2D9A4}} \color[HTML]{000000} 17.5 & {\cellcolor[HTML]{CAEA9E}} \color[HTML]{000000} 37.3 & & {\cellcolor[HTML]{BCE4A0}} \color[HTML]{000000} 30.3 & {\cellcolor[HTML]{EBF7A0}} \color[HTML]{000000} 56.7 \\
    & 3 & 
{\cellcolor[HTML]{9FD8A4}} \color[HTML]{000000} 16.1 & {\cellcolor[HTML]{FBA35C}} \color[HTML]{000000} 144.7 & & {\cellcolor[HTML]{A7DBA4}} \color[HTML]{000000} 19.8 & {\cellcolor[HTML]{C6E89F}} \color[HTML]{000000} 34.7 & & {\cellcolor[HTML]{8FD2A4}} \color[HTML]{000000} 9.1 & {\cellcolor[HTML]{ACDDA4}} \color[HTML]{000000} 22.2 & & {\cellcolor[HTML]{B3E0A2}} \color[HTML]{000000} 25.4 & {\cellcolor[HTML]{E8F69B}} \color[HTML]{000000} 53.8 \\
    & 10 & 
{\cellcolor[HTML]{97D5A4}} \color[HTML]{000000} 13.1 & {\cellcolor[HTML]{FDAD60}} \color[HTML]{000000} 141.3 & & {\cellcolor[HTML]{9FD8A4}} \color[HTML]{000000} 16.3 & {\cellcolor[HTML]{BCE4A0}} \color[HTML]{000000} 30.6 & & {\cellcolor[HTML]{89D0A4}} \color[HTML]{000000} 7.3 & {\cellcolor[HTML]{9CD7A4}} \color[HTML]{000000} 15.4 & & {\cellcolor[HTML]{A7DBA4}} \color[HTML]{000000} 19.1 & {\cellcolor[HTML]{D8EF9B}} \color[HTML]{000000} 43.9
    \end{tabular}
    \label{tab: eigenvalue_errors}
\end{table*}

The ridge-regression model outperformed the oneshot, average, and sH models for all systems, both for the screening parameters and the eigenenergies. The mean absolute error of the eigenenergies of the ridge-regression model is below $\SI{25}{meV}$ for all test systems already after 1 training snapshot. The oneshot model behaves poorly for water with an average error over four times that of the ridge-regression model, while for \ch{CsSnI3} it performs much better, aided by the fact that the variational orbitals fall into groups of orbitals of the same character that do not change from one snapshot to the next. That said, that the oneshot model does not capture any of the trends of the screening parameters between orbitals of the same character (as can be seen in the scatter plot of the screening parameters, in which each cluster of points show no internal correlation). Meanwhile, the average and self-Hartree models predict eigenenergies with an average error above $\SI{40}{meV}$ for water and even above $\SI{200}{meV}$ for \ch{CsSnI3}. The fact that the 1-snapshot ridge-regression model outperforms the oneshot model in both cases demonstrates that even if one only calculates screening parameters for a single snapshot, it is better to construct a ridge regression model from this data rather than using the computed screening parameters directly for other snapshots.

In most applications, one is interested in specific orbital energies and not in quantities averaged over all eigenenergies. Therefore, it is important for the error distribution of the eigenenergies not to have a long tail. In this regard, the ridge-regression model also performs very well. After three training snapshots, ridge regression predicts no single eigenenergy with an error larger than $\SI{150}{meV}$. In comparison, the average model and the sH model predict many eigenenergies with an error larger than $\SI{200}{meV}$; for \ch{CsSnI3} many have errors larger than even $\SI{500}{meV}$. 

The most important eigenenergies for many applications are the highest occupied molecular orbital (HOMO) and the lowest unoccupied molecular orbital (LUMO) energies, or in bulk systems (such as $\ch{CsSnI3}$) the valence band maximum (VBM) and the conduction band minimum (CBM); in this work, we will treat these terms synonymously. Another important quantity is the band gap (the difference between the VBM and CBM). While the error of the CBM is below the average error in all cases, the error in the VBM is above-average in all cases. This is not a fault of the models but suggests that the screening parameters have a larger influence on the VBM than the CBM. Even still, for most applications the VBM is predicted sufficiently accurately by the ridge regression after 2 or 3 training snapshots. Why the self-Hartree model fails for these two systems after showing promise in the preliminary study of acenes is analyzed in Supplementary Discussion~1.

\begin{figure}[t]
    \centering
\includegraphics[width=\onecolwidth]{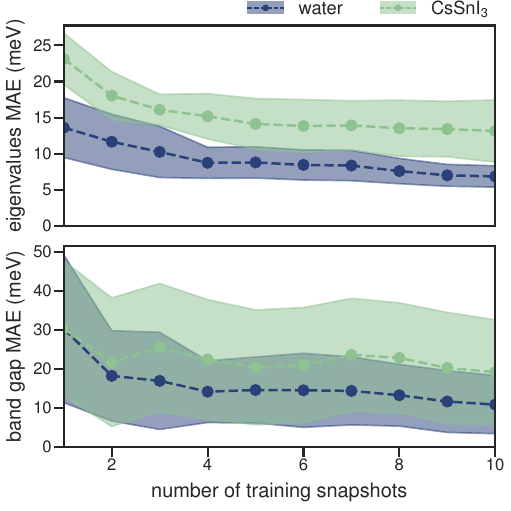}
\caption{\textbf{Convergence of the results as a function of the number of snapshots used for training.} The top panel shows the mean absolute error in the eigenvalues, while the bottom panel shows the band gap. To get a statistical average, both quantities were obtained for 10 test snapshots. The dotted line indicates the mean, while the shaded region the standard deviation across the 10 test snapshots.}\label{fig: convergence analysis}
\end{figure}

Finally, we examine the convergence of the ridge-regression model with respect to the number of training snapshots in greater detail. \Cref{fig: convergence analysis} shows the convergence of the eigenvalues and the band gap as a function of the number of training snapshots. As we have already seen, the error of the HOMO energy and, correspondingly, the error of the band gap is larger than the mean absolute error. Nevertheless, both quantities become small (i.e. less than 20 meV) already after a few (2 to 4) training snapshots. To put this in perspective, Koopmans functionals typically predict orbital energies and band gaps within \SI{200}{meV} of experiment \cite{Nguyen2018,Colonna2019} i.e. the error introduced by the ML model is acceptably small, and the accuracy of the predicted band gaps remains state-of-the-art.

\subsection{Speed-up}
The main goal in developing the ML model is to speed up the calculations with Koopmans spectral functionals while maintaining high accuracy. In the preceding section, we saw that the model achieves a satisfactory level of accuracy after a few training snapshots. In this section, we turn to look at the corresponding speedups.

\begin{figure}[t]
    \centering
\includegraphics[width=\onecolwidth]{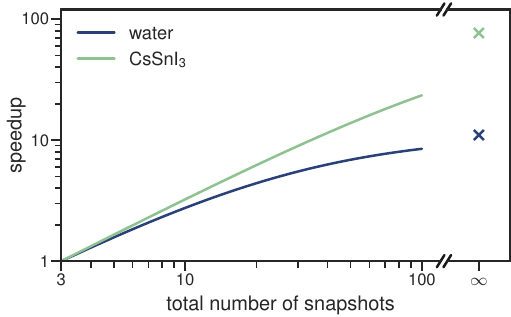}
\caption{\textbf{Speed-ups when using the ridge-regression model to predict screening parameters.} The timings show the speed-up depending on the total number of snapshots for which Koopmans functional calculations must be performed, assuming three snapshots are required for training and the remaining snapshots use the ridge-regression model to predict the screening parameters. All calculations were performed on an Intel Xeon Gold 6248R machine with 48 cores.}\label{fig: speedup}
\end{figure}

We find that the ratio of the time required for an entire Koopmans calculation when (a) all screening parameters are computed ab initio relative to when (b) all screening parameters are predicted using the ML model is 80 for \ch{CsSnI3} and 11 for water. However, this does not factor in the cost of training the model in the first place. \Cref{fig: speedup} shows the anticipated speed-ups for performing Koopmans calculations on a given number of snapshots assuming that training requires three snapshots for training (and therefore three snapshots performed \emph{ab initio}). For performing Koopmans calculations on 20 snapshots one obtains a speedup of roughly 4.4 for the water system and of roughly 6.2 for the \ch{CsSnI3} system. Because training the model is a one-off cost, for infinitely many configurations we approach the aforementioned 80-fold and 11-fold speedups.

We note that there are scenarios in which the ML model will provide even greater speed-ups than these. For example, the model could be trained on a small supercell and then employed for calculations on larger supercells. Because the cost of DFT calculations scales cubically with system size (\emph{cf.} linearly with the number of snapshots), machine-learning promises even greater speed-ups for such schemes. Note that this approach is only possible because the descriptors are spatially localized and therefore invariant with respect to the system size.

\subsection{Transferability}

Thus far the ridge-regression models have only been applied to the same system that they were trained on. In this section, we seek to quantify the models' transferability. Due to the simplicity of the ridge-regression model and the small amount of training data, we expect these models to not be very transferable.

The transferability of the ridge-regression model trained on 10 snapshots of liquid water was tested on two other phases of water: ice XI and isolated water molecules. The ice XI geometries correspond to a crystalline hydrogen-bonding network of water molecules, with atomic positions displaced from the pristine crystal structure using the self-consistent harmonic approximation in order to accurately reflect the quantum zero-point motion of the atoms at \SI{0}{K} (and to thereby introduce some inhomogeneity into the data for the model to try and capture) \cite{Monacelli2018}. Meanwhile, the test set of isolated water molecules correspond to 200 individual molecules that were extracted from the liquid water snapshots reserved for testing. More details about the ice XI and isolated water molecule structures can be found in the Methods section below and Supplementary Figure~6.

\begin{figure}
   \centering
   \includegraphics[width=\onecolwidth]{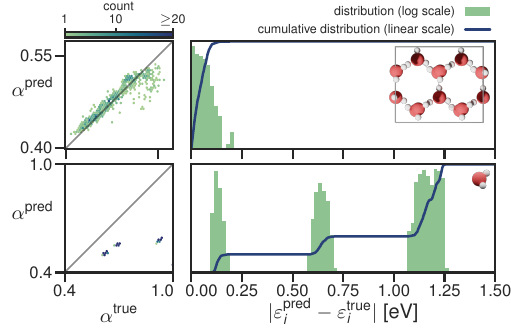}
   \caption{\textbf{Transferability of the ridge-regression model trained on 10 water snapshots.} The top panel shows the accuracy of the model when applied to ice XI, and the bottom panel shows the accuracy of the model when applied to isolated water molecules. Refer back to \cref{fig: water_cls} for further explanation. Note that the scale of the $x$ axis is three times larger than that of \cref{fig: water_cls,fig: CsSnI3}.}
   \label{fig:water transferability}
\end{figure}

The results of applying the model trained on liquid water to these two test systems are shown in \cref{fig:water transferability}. The model performs relatively well for ice XI: the electrons in ice XI still find themselves in a system of hydrogen-bonded water molecules with local atomic environments and a level of electronic screening not dissimilar to liquid water. This is not the case, however, for the isolated water molecules, for which the model catastrophically fails. This is expected, because the electronic screening is totally different for isolated water molecules (where there are no surrounding electrons to screen the formation of a charged state when an electron is removed from/added to the system) than in liquid water (where the surrounding electrons screen the formation of such a charged state).

\begin{figure}
   \centering
   \includegraphics[width=\onecolwidth]{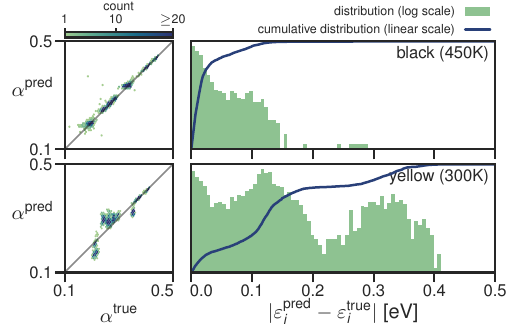}
   \caption{\textbf{Transferability of the ridge-regression model trained on 10 \ch{CsSnI3} snapshots at 250K.} The top panel shows the accuracy of the model when applied to snapshots at 450K, and the bottom panel shows the accuracy of the model when applied to the yellow phase at 300K. Refer back to \cref{fig: water_cls} for further explanation.}
   \label{fig:cssni3 transferability}
\end{figure}

The transferability of the ridge-regression model for \ch{CsSnI3} was also tested against two test systems. In this case, the two systems corresponded to (a) the same phase, but with atomic replacements corresponding to \SI{450}{K} (\emph{cf.} the systems on which the model was trained, which correspond to \SI{250}{K}) and (b) the yellow phase of \ch{CsSnI3}. This second phase contains local atomic environments that qualitatively differ from those found in the black phase of \ch{CsSnI3} -- for example, the caesium atoms go from a twelve-fold to a nine-fold coordination environment with much shorter bond lengths. For more details, see the Methods and Supplementary Figure~7. As shown in \cref{fig:cssni3 transferability} the model predicts the screening parameters for the 450K phase well with only a handful of outliers, but the model does not transfer well to the novel atomic environments found in the yellow phase.

These two tests demonstrate that for the model to be transferable, the macroscopic screening ought to be similar (as was not the case for isolated water molecules) and the atomic environments should be familiar (as was not the case for the yellow phase of \ch{CsSnI3}). If one wanted a machine-learning model to be more transferable, more training data and/or a more sophisticated ML model would be required. It would also be possible to employ ensemble and/or active learning approaches to detect and correct failures \cite{Settles2009,Sagi2018}. Ultimately, the choice of machine-learning model needs to strike a balance between transferability and the cost of training, which will differ from one application to the next.

\section{Discussion}
This work presents a machine learning framework to predict electronic screening parameters via ridge regression performed on translationally- and rotationally-invariant power spectrum descriptors of orbital densities. This framework is able to predict the screening parameters for Koopmans spectral functionals with sufficient accuracy and with sufficiently little training data so as to dramatically decrease the computational cost of these calculations while maintaining their state-of-the-art accuracy, as demonstrated on the test cases of liquid water and the halide perovskite \ch{CsSnI3}.

It is somewhat surprising that it is possible to accurately predict screening parameters directly from orbital densities, because the screening parameters are not explicitly determined by orbital densities themselves but instead are related to the \emph{response} of these densities (refer back to \cref{eq: alpha_LR}). Nevertheless, for the systems studied in this work the relationship between the orbital's density (plus the surrounding electron density) and its screening parameter learned by a ridge-regression model gives very accurate predictions of the screening parameters for orbitals it has never seen before --- albeit for the same system; we do not expect the model to be transferable across different systems.

Work is already ongoing to use this machine-learning framework to predict the temperature-dependent spectral properties of materials of scientific interest.

In this work, the strategy of learning the mapping from orbital densities to screening parameters has been tailored for use with Koopmans functionals. However, the potential applications of this method are not limited to these functionals. For methods that are related to Koopmans functionals (as introduced earlier) the applicability of this strategy varies. For example, the localized orbital scaling correction has a matrix $\kappa_{i j}$ ($i$ and $j$ being the indices of localized orbitals), whose elements are calculated in a very similar manner to Koopmans screening parameters \cite{Li2018,Mahler2022a}. In this instance, the present machine-learning framework could be applied virtually unchanged (and would be even more beneficial, given that there are now $\mathcal{O}(N_\text{orb}^2)$ parameters to compute). In contrast, this framework would be less useful in the context of optimally-tuned hybrid functionals \cite{Stein2010,Kronik2012,Wing2021}, because only a solitary parameter (the range separation parameter) is determined via a piecewise linearity criterion. More broadly, if one wants to construct a model that can predict energy curvatures of the form $\frac{d^2 E}{d f^2}$ for some generic orbital occupation $f$, then this work demonstrates that it is efficient to instead construct its frozen-orbital counterpart $\frac{\partial^2E}{\partial f^2}$ and then learn the mapping from this quantity to the screened quantity $\frac{d^2E}{d f^2}$.

%

\section{Methods}\label{sec:methods}

All of the Koopmans functional calculations presented in this work were performed using \texttt{Quantum ESPRESSO} \cite{Giannozzi2009, Giannozzi2017} and the \texttt{koopmans} package \cite{Linscott2023}. The most relevant details of these calculations can be found below, while full input and output files can be found in the \href{https://doi.org/10.24435/materialscloud:w1-ev}{Materials Cloud Archive} \cite{Schubert2024}.

\subsection{Liquid water}
The liquid water system studied in this work is a simple cube with a side length of $\SI{9.81}{\angstrom}$ containing 32 water molecules. 20 snapshots were taken from \emph{ab initio} molecular dynamics trajectories at \SI{300}{K} with the nuclei treated classically, as presented in Ref.~\citenum{Chen2016} (and as previously studied using Koopmans functionals in Ref.~\citenum{deAlmeida2021}). Those molecular dynamics calculations used the revised Vydrov and Van Voorhis (rVV10) van der Waals exchange-correlation functional \cite{Vydrov2010}, a kinetic energy cutoff of \SI{85}{Ry}, and SG15 optimized norm-conserving Vanderbilt pseudopotentials \cite{Hamann2013}. The 20 snapshots on which we then performed Koopmans calculations were taken from the \SI{10}{ps} production run, each \SI{0.5}{ps} apart to ensure that the snapshots were not autocorrelated. (While this information is worth repeating, note that the details of how the snapshots were generated ultimately have little relevance to the task of predicting the screening parameters.)

The Koopmans functional calculations on water used PBE as a base functional \cite{Perdew1996}, an energy cutoff of $80\text{ Ry}$ (selected as the result of a convergence analysis), and the Makov-Payne periodic image correction scheme \cite{Makov1995}. The system is spin-unpolarized and therefore the 32 water molecules give rise to 128 occupied orbitals. We also included 64 empty orbitals (which are typically included in Koopmans functional calculations in order to be able to be able to describe excitations involving electron addition i.e. photoabsorption). As such, each snapshot represents 192 datapoints (i.e. 192 pairs of orbital density descriptors and the corresponding screening parameter calculated \emph{ab initio}), with the entire dataset corresponding to 3840 datapoints. 

\subsection{Ice XI}
For the purposes of testing the transferability of the ridge-regression model, it was tested on two further water systems. The first of these was ice XI, an ordered phase of ice that forms below \SI{72}{K} in the presence of a small amount of alkali-metal hydroxide \cite{Tajima1982}. The structure of ice XI is shown in Supplementary Figure~6. The atomic positions used in this study correspond to a $T = 0$\,K calculation with the self-consistent harmonic approximation, as reported in Ref.~\cite{Monacelli2018}. In this scheme, the atoms are displaced from their equilibrium positions in a specific pattern to accurately capture the effects of zero-point quantum fluctuations.

The calculations were performed on a $2 \times 2 \times 2$ supercell of the eight-molecule, \SI{4.39}{\AA}-by-\SI{7.59}{\AA}-by-\SI{7.59}{\AA} orthorhombic cell. The supercell therefore contains 288 occupied orbitals and 192 empty orbitals in total. The simulation cell aside, the calculations used the same settings as those of liquid water. 

\subsection{Isolated water molecules}
The second phase of water that was used to test the transferability of the ridge-regression model was isolated water molecules. In order to generate a set of non-identical water molecules with varying geometries, 200 water molecules were taken from the liquid water snapshots and considered in isolation, placed in a cubic cell with length 15\,\AA. These calculations used the same settings as those of liquid water but with the macroscopic dielectric constant (used in the Makov-Payne correction) set to 1.

\subsection{The black $\alpha$ phase of \ch{CsSnI3}} \label{sec: comp. CsSnI3}
The perovskite system studied in this work was a 2$\times$2$\times$2 supercell of the 5-atom primitive cell of the black $\alpha$ phase of \ch{CsSnI3}; a cubic cell with a side length of $\SI{12.35}{\angstrom}$. The snapshots used in this work were generated via the stochastic self-consistent harmonic approximation \cite{Monacelli2021}. This method simulates the thermodynamic properties originating from the quantum and thermal anharmonic motion of the ions \cite{Monacelli2021}, where the ionic energy landscape is stochastically sampled and then evaluated within DFT. Unless otherwise noted, the configurations correspond to \SI{250}{K} \cite{Monacelli2023}.

The Koopmans functional calculations on \ch{CsSnI3} used a kinetic energy cutoff at $70\text{ Ry}$ (determined as the result of a convergence analysis), norm-conserving pseudopotentials from the Pseudo-Dojo library \cite{VanSetten2018}, PBEsol as the base exchange-correlation functional \cite{Perdew2008}, and Makov-Payne periodic image corrections \cite{Makov1995}. The system is spin-unpolarized and has 176 occupied bands, and 24 empty bands were included \emph{i.e.} each snapshot represents 200 datapoints, with the entire dataset corresponding to 4000 datapoints.

\subsubsection{The yellow phase of \ch{CsSnI3}}
The second ``yellow" phase of \ch{CsSnI3} was used to assess the transferability of the ridge-regression model. This phase appears to be the ground state of \ch{CsSnI3}, with the black $\alpha$ phase metastable under ambient conditions \cite{Monacelli2023}. The calculations on the yellow phase were performed on a 2$\times$2$\times$2 supercell of the 5-atom orthorhombic primitive cell, leading to a \SI{10.49}{\angstrom}-by-\SI{9.49}{\angstrom}-by-\SI{17.77}{\angstrom} simulation cell containing 40 atoms. The atoms were displaced according to the stochastic self-consistent harmonic approximation to correspond to nuclear motion at \SI{300}{K}. Otherwise, the calculations used the same settings as those of the black phase.

The atomic environments found in black and yellow \ch{CsSnI3} at different temperatures are compared in Supplementary Figure~7.

\subsection{The ridge-regression model}
The ridge-regression model was implemented using the \texttt{scikit-learn} library \cite{Pedregosa2011}. The descriptor hyperparameters were set to $n_\mathrm{max}=6$, $l_\mathrm{max}=6$, $r_\mathrm{min}=\SI{0.5}{a_0}$, and $r_\mathrm{max}=\SI{4.0}{a_0}$. An exhaustive grid search over the hyperparameters (with all possible combinations of $n_\mathrm{max} \in \{1, 2, 4, 8\}$, $l_\mathrm{max} \in \{1, 2, 4, 8\}$, $r_\mathrm{min}/a_0 \in \{0.5, 1, 2\}$, and $r_\mathrm{max}/a_0 \in \{2, 4, 8\}$) revealed that for $4\leq n_\mathrm{max}\leq 8$, $4\leq l_\mathrm{max}\leq 8$, and $0.5 \leq r_\mathrm{min}\leq 1$, the mean absolute error in the predicted screening parameters was not very sensitive to the choice of hyperparameters. For the ridge-regression model, the regularization parameter was set to 1 as the result of 10-fold cross-validation, and the input vectors $p^i$ were standardized.

\backmatter

\bmhead{Data availability}
The data underpinning this work can be downloaded from the \href{https://doi.org/10.24435/materialscloud:4s-xf}{Materials Cloud Archive} \cite{Schubert2024}.

\bmhead{Code availability}
The \texttt{koopmans} code used to generate the results of this paper is open-source and is available at \href{https://github.com/epfl-theos/koopmans/}{github.com/epfl-theos/koopmans}. \href{https://github.com/epfl-theos/koopmans/releases/tag/v1.0.0}{Versions 1.0.0} and  \href{https://github.com/epfl-theos/koopmans/releases/tag/v1.1.0}{1.1.0} of the code were used.

\bmhead{Acknowledgements}
The authors thank Nicola Colonna, Lorenzo Monacelli, and Martin Uhrin for helpful discussions. EL gratefully acknowledges financial support from the Swiss National Science Foundation (grant numbers 179138 and 213082). This research was also supported by the NCCR MARVEL, a National Centre of Competence in Research, funded by the Swiss National Science Foundation (grant number 205602).

\bmhead{Author contributions}
YS: Methodology, Software, Validation, Investigation, Writing -- Original Draft. SL:~Supervision, Writing -- Review \& Editing. NM:~Conceptualization, Supervision, Writing -- Review \& Editing. EL: Conceptualization, Methodology, Software, Investigation, Supervision, Writing -- Review \& Editing

\bmhead{Competing interests}
The authors declare no competing interests.

\bibliography{references}


\end{document}


\title{Supplementary information: Predicting electronic screening for fast Koopmans spectral functional calculations}

\author[1]{\fnm{Yannick} \sur{Schubert}\orcidlink{0000-0002-9676-9134}}

\author[1]{\fnm{Sandra} \sur{Luber}\orcidlink{0000-0002-6203-9379}}

\author[2,3]{\fnm{Nicola} \sur{Marzari}\orcidlink{0000-0002-9764-0199}}

\author*[3,4]{\fnm{Edward} \sur{Linscott}\orcidlink{0000-0002-4967-9873}\email{edward.linscott@psi.ch}}

\affil[1]{\orgdiv{Department of Chemistry}, \orgname{University of Zurich}, \orgaddress{\postcode{8057} \city{Zurich}, \country{Switzerland}}}

\affil[2]{\orgdiv{Theory and Simulations of Materials (THEOS) and National Centre for Computational Design and Discovery of Novel Materials (MARVEL)}, \orgname{École Polytechnique Fédérale de Lausanne}, \orgaddress{\postcode{1015} \city{Lausanne}, \country{Switzerland}}}

\affil[3]{\orgdiv{Center for Scientific Computing, Theory and Data}, \orgname{Paul Scherrer Institute}, \orgaddress{\postcode{5352} \city{Villigen PSI}, \country{Switzerland}}}

\affil[4]{\orgdiv{National Centre for Computational Design and Discovery of Novel Materials (MARVEL)}, \orgname{Paul Scherrer Institute}, \orgaddress{\postcode{5352} \city{Villigen PSI}, \country{Switzerland}}}


\maketitle

\renewcommand{\figurename}{Supplementary Figure}
\renewcommand{\tablename}{Supplementary Table}

\section{Correlation between self-Hartree energies and screening parameters}
\subsection{Acenes}
\label{sec: acenes}
\begin{figure}[h]
    \centering
    \includegraphics[width=0.5\columnwidth]{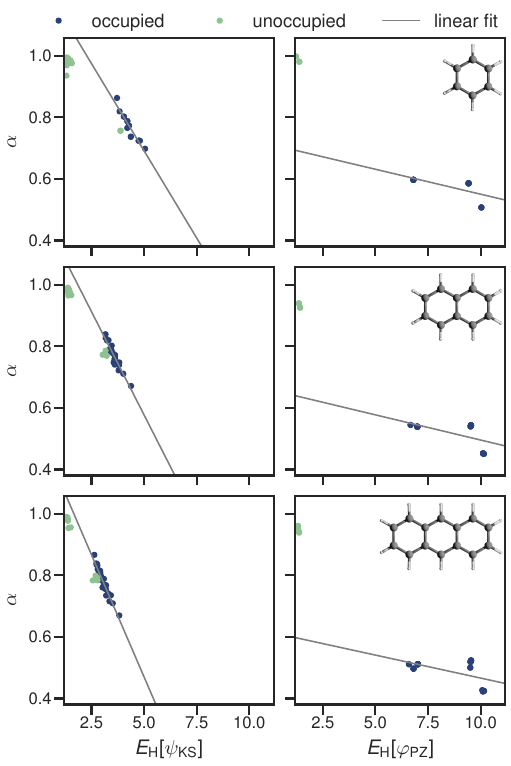}
    \caption{\textbf{Correlation between the self-Hartree energies and the screening parameters for three different acenes:\ benzene, naphthalene, and anthracene.} The left panel shows the results with Kohn-Sham initial orbitals and the right panel the results with Perdew-Zunger initial orbitals. Green dots correspond to filled and red dots to empty orbitals. The linear interpolation (black line) was performed using only the occupied orbitals.}\label{fig: sH acenes}
\end{figure}

Correlations exist between the self-Hartree energies and screening parameters for different acenes (benzene, naphthalene, and anthracene), as shown in \cref{fig: sH acenes}. In contrast to the liquid water and perovskite studied in the main text, these are non-periodic systems and correspondingly instead of Wannier orbitals the variational orbitals are chosen as either Kohn-Sham or Perdew-Zunger orbitals. The former will be delocalized across the molecule; the latter will be more localized. The correlations are stronger in the case of the Kohn-Sham orbitals.

\subsection{Water}

\begin{figure}[h]
    \centering
\includegraphics[width=0.5\columnwidth]{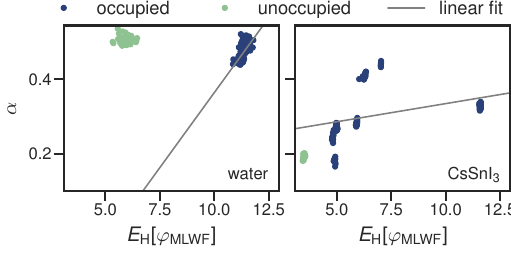}
\caption{\textbf{Correlation between the self-Hartree energies and the screening parameters of the MLWFs of water and \ch{CsSnI3}.} The line is fitted to the occupied orbital data.}\label{fig: sH test systems}
\end{figure}

Testing the same correlation for the water system, we find that the correlation is much weaker (see \cref{fig: sH test systems}). For water, the empty states and the occupied states have very similar screening parameters but very different self-Hartree energies. This suggests that the self-Hartree energies alone don't contain sufficient information to predict the screening parameters. If we had predicted occupied and empty states with the same sH model instead of two separate ones, we would have obtained similarly poor results as for the mean model.

\subsection{CsSnI\textsubscript{3}}
For \ch{CsSnI3}, there is almost no correlation between the self-Hartree energies and the screening parameters of MLWFs (see \cref{fig: sH test systems}). In the individual clusters of data, there are many orbitals with very similar self-Hartree energies but different screening parameters. Moreover, the clusters as a whole seem to show little correlation with the self-Hartree energies.

\section{Basis functions details}
For the basis functions used to represent orbital densities, we adapt the choice (following Ref.~\citenum{Himanen2020} and many others) of taking real-valued spherical harmonics as angular basis functions and Gaussian basis functions as radial basis functions.

\subsection{Angular basis functions}
The real-valued spherical harmonics are given by
\begin{align*}
    Y_{lm}(\theta,\varphi)=\begin{cases} 
    \sqrt{2} (-1)^m \Im\left[Y_l^{|m|}(\theta,\varphi)\right] & \text{if } m<0\\
    Y_l^0 & \text{if } m=0\\
    \sqrt{2}(-1)^m \Re\left[Y_l^m(\theta,\varphi)]\right] & \text{if } m>0
    \end{cases}
\end{align*}
and define an orthogonal and complete set of angular basis functions. Here, $Y_l^m$ are the complex orthonormalized spherical harmonics:
\begin{align}
    Y_l^m(\theta,\varphi)=\sqrt{\frac{(2l+1)}{4\pi}\frac{(l-m)!}{(l+m)!}}P_l^m(\cos(\theta))e^{i m\varphi}
\end{align}
and $P_l^m$ are the associated Legendre polynomials. The expansion into spherical harmonics are truncated after some maximum value $l_\mathrm{max}$.

\subsection{Radial basis functions}
For the radial basis functions, we construct a set of orthonormal basis functions
\begin{align*}
    g_{nl}(r)=\sum\limits_{n'=1}^{n_\mathrm{max}}\beta_{nn'l}\phi_{n'l}(r) 
\end{align*}
out of a set of linearly independent Gaussians:
\begin{align*}
    \phi_{nl}(r)=r^l e^{-\gamma_{nl} r^2}.
\end{align*}
The decay parameters $\gamma_{nl}$ are chosen such that $\phi_{nl}$ decays to a threshold value of $10^{-3}$ for cutoff radii taken on an evenly spaced grid from $r_\mathrm{min}$ to $r_\mathrm{max}$. This means for each $n\in\{1,\dots,n_\mathrm{max}\}$ the cutoff radius is given by
\begin{align*}
    r_{\mathrm{thr},n} &= r_\mathrm{min} + \frac{n-1}{n_\mathrm{max}-1} (r_\mathrm{max}-r_\mathrm{min})
\end{align*}
The coefficients $\beta_{nn'l}$ are obtained with a Löwdin orthogonalization procedure \cite{Lowdin1950}: $\boldsymbol{\beta}_l=\textbf{S}_l^{-1/2}$, where
\begin{align}
    {(\boldsymbol{S}_l)}_{nn'}=\langle \phi_{nl}|\phi_{n'l}\rangle = \int\limits_0^\infty dr \, r^{2(l + 1)} e^{-(\gamma_{nl} + \gamma_{n'l})r^2},
\end{align}
\subsection{Reconstructed orbital densities}

\begin{figure}[b]
    \centering
    \begin{subfigure}{0.23\textwidth}
    \includegraphics[width=\textwidth,trim=90 90 90 90, clip]{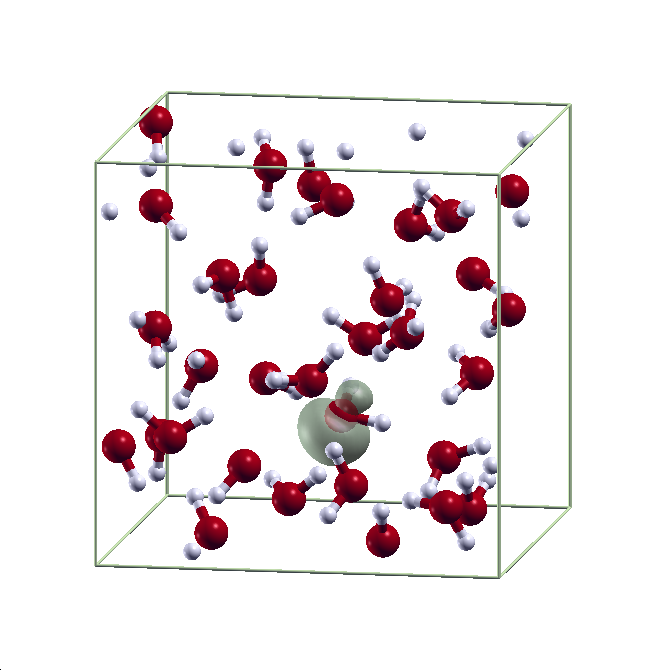}
    \caption{original}
\end{subfigure}
\begin{subfigure}{0.23\textwidth}
    \includegraphics[width=\textwidth,trim=90 90 90 90, clip]{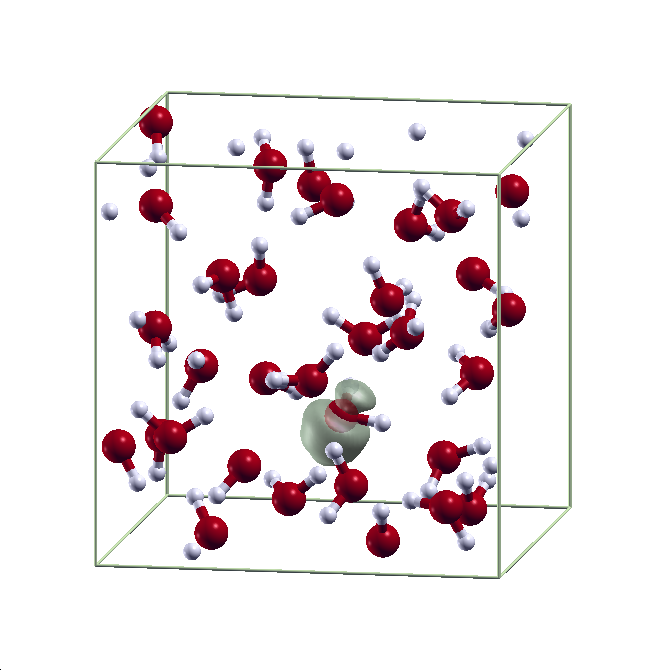}
    \caption{reconstructed}
\end{subfigure}
\caption{\textbf{Original and reconstructed occupied $sp^3$ Wannier function centered on an oxygen atom in the water system.} The isosurfaces are plotted at $0.005 a_0^{-3}$.}\label{fig: water_orbitals}
\end{figure}

\begin{figure}[t]
    \centering
\begin{subfigure}{0.23\textwidth}
    \includegraphics[width=\textwidth,trim=90 90 90 90, clip]{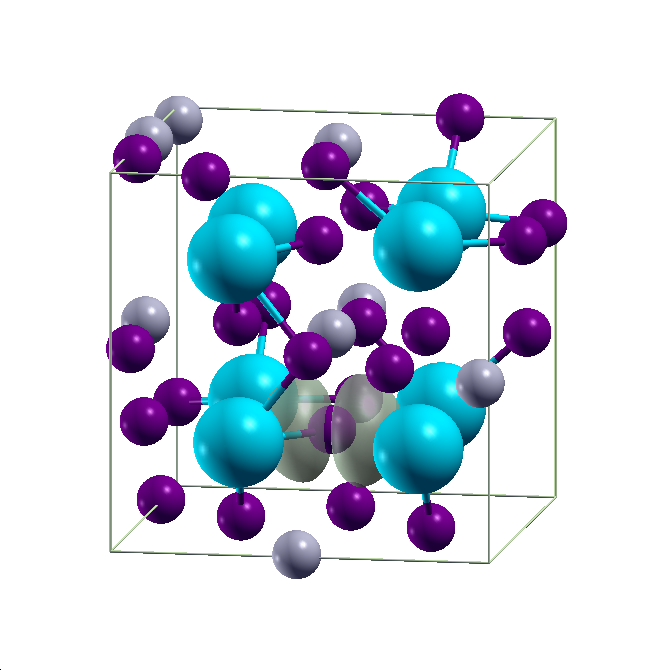}
    \caption{original}
\end{subfigure}
\begin{subfigure}{0.23\textwidth}
    \includegraphics[width=\textwidth,trim=90 90 90 90, clip]{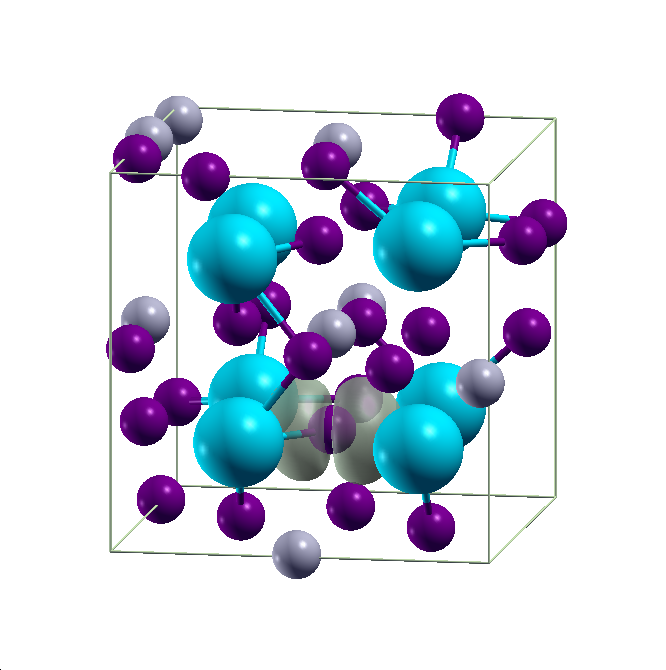}
    \caption{reconstructed}
\end{subfigure}
\caption{\textbf{Original and reconstructed occupied $p$ Wannier orbital centered on an iodine atom in \ch{CsSnI3}.} The isosurfaces are plotted at $0.0005 a_0^{-3}$.}\label{fig: CsSnI3 orbitals}
\end{figure}

In \cref{fig: water_orbitals,fig: CsSnI3 orbitals} we demonstrate the efficacy of the orbital density descriptors by comparing variational orbital densities against reconstructed densities i.e. densities obtained by performing the decomposition into a truncated basis set as described above and in the main text, and then taking the linear combination of the basis functions with the obtained expansion coefficients. Without the truncation, the original and reconstructed densities should be identical.

\section{Sensitivity of the eigenvalues to the screening parameters}
In the main text it was shown that the Koopmans correction shifts DFT Kohn-Sham eigenvalues by
\begin{align}
    \Delta \varepsilon_i = \varepsilon^\mathrm{KI}_i - \varepsilon^\mathrm{DFT}_i
    =
    \sum_{jk} \alpha_j U_{ij}U_{ki}^\dag\braopket{\varphi_k}{\hat v_j^\mathrm{KI}}{\varphi_j}
\end{align}
where the variational and canonical orbitals are related via a unitary rotation (i.e. $\ket{\psi_i} = \sum_j U_{ij} \ket{\varphi_j})$, and for occupied orbitals the shift simplifies to
\begin{align}
    \Delta \varepsilon_{i \in \mathrm{occ}} = 
    \sum_{j} \alpha_j U_{ij}U_{ji}^\dag
    \left(-E_\mathrm{Hxc}[\rho - \rho_j]+E_\mathrm{Hxc}[\rho - \rho_j + n_j] - \int d\mathbf{r} \, v_\mathrm{Hxc}[\rho - \rho_j + n_j](\mathbf{r})  n_j(\mathbf{r}) \right)
\end{align}
For a concrete example, calculations were performed on a single snapshot of liquid water with the screening parameters of all orbitals increased from 0 (i.e. the DFT solution) to 1 (the fully unscreened limit). The resulting change in the quasiparticle energies of the system are shown in Supplementary Figure~\ref{fig: alpha scan}. Note that in calculations where the screening parameter is calculated \emph{ab initio}, each variational orbital has its own screening parameter in the range of 0.43 to 0.53. The eigenvalues change by approximately 1 eV across this window.
\begin{figure}[t]
    \centering
    \includegraphics[width=0.5\columnwidth]{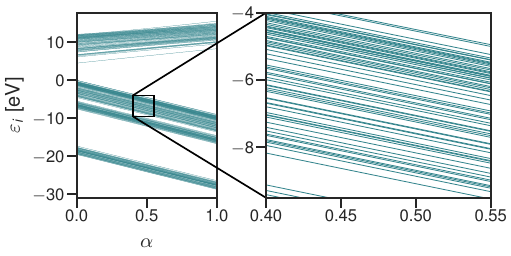}
    \caption{\textbf{The KI@PBE eigenvalues of a liquid water system as a function of the screening parameter.} The same screening parameter was used for every variational orbital in the system. The eigenvalues energies are provided relative to the HOMO energy of the DFT solution. The inset shows the quasi-particle energies of the uppermost occupied states across the range of values measured \emph{ab initio}.} \label{fig: alpha scan}
\end{figure}

\begin{figure}[t]
    \centering
    \begin{minipage}{0.275\textwidth}
        \begin{tikzpicture}
            \node[anchor=south west, inner sep=0] (image) at (0, 0) {\includegraphics[width=\textwidth]{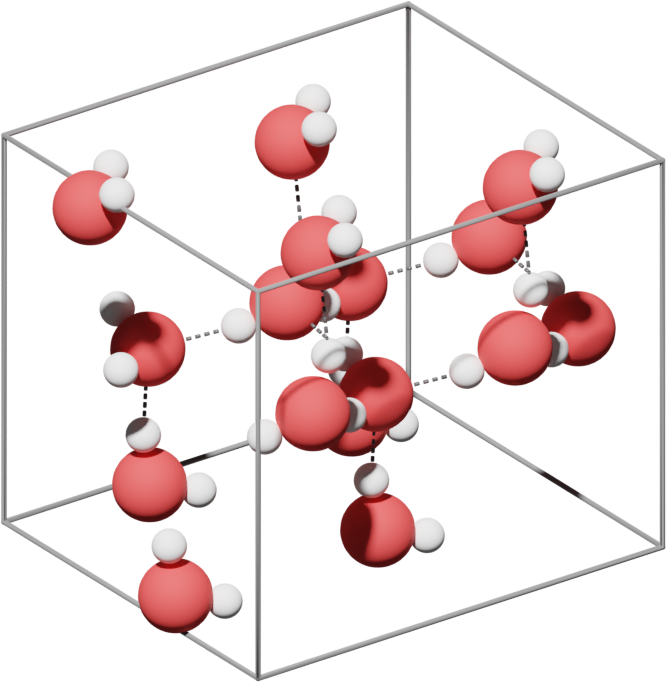}};
            \node[anchor=north west, text=black, inner sep=0,font=\bfseries\sf] at (image.north west) {a\vphantom{abc}};
        \end{tikzpicture}
    \end{minipage}%
    \hspace{0.02\textwidth}
    \begin{minipage}{0.175\textwidth}
        \begin{tikzpicture}
            \node[anchor=south west, inner sep=0] (image) at (0, 0) {\includegraphics[width=\textwidth]{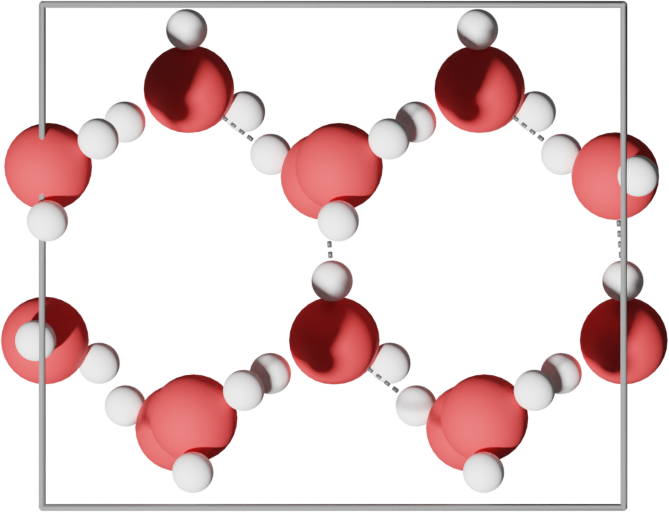}};
            \node[anchor=north west, text=black, inner sep=2pt,font=\bfseries\sf, xshift=-10pt] at (image.north west) {b\vphantom{abc}};
        \end{tikzpicture}

        \vspace{0.02\textwidth}
        \begin{tikzpicture}
            \node[anchor=south west, inner sep=0] (image) at (0, 0) {\includegraphics[width=\textwidth]{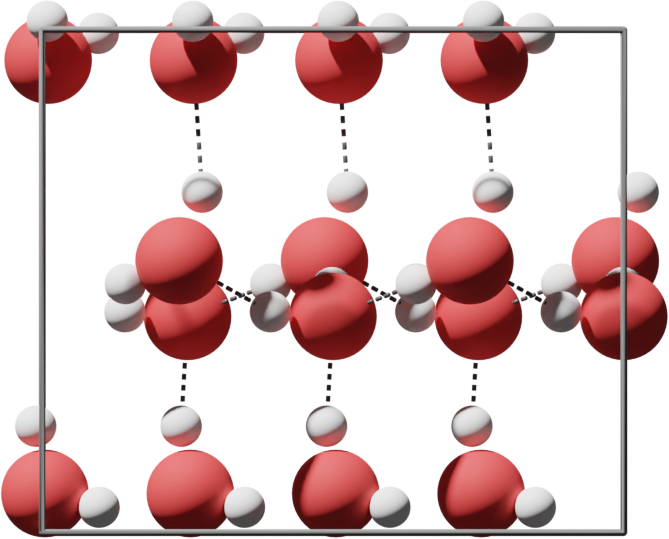}};
            \node[anchor=north west, text=black, inner sep=2pt,font=\bfseries\sf, xshift=-10pt] at (image.north west) {c\vphantom{abc}};
        \end{tikzpicture}
    \end{minipage}
    \caption{\textbf{The crystal structure of ice XI.} (a) A $2 \times 1 \times 1$ supercell of the eight-molecule orthorhombic cell. The smaller insets show the view (b) from above and (c) from the side.}
    \label{fig: ice}
\end{figure}

\begin{figure}[t]
    \centering
    \includegraphics[width=0.5\textwidth]{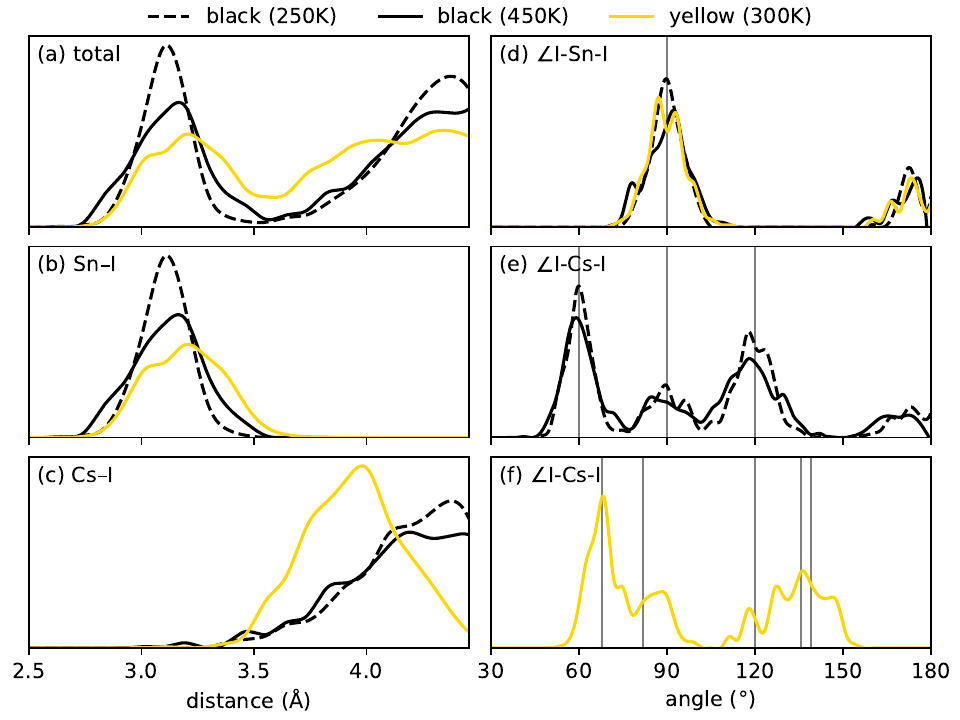}
    \caption{\textbf{Comparison of coordination environments of \ch{CsSnI3} for the three systems studied.} (a-c) show radial distribution functions; (d-f) show bond angle distributions. The grey vertical lines show the bond angles for pristine octahedral (six-fold), cuboctahedral (twelve-fold), and tricapped trigonal prismatic (nine-fold) coordination.}
    \label{fig: rdfs}
\end{figure}

\renewcommand{\refname}{Supplementary References}
\bibliography{references}